\documentclass[preprint,12pt]{elsarticle}

\usepackage{amsmath,amssymb}
\usepackage{bm}
\usepackage{multirow}
\usepackage{subfigure}
\usepackage{lineno}
\biboptions{sort&compress}
\usepackage{tikz}
\usepackage{makecell}

\newcommand\undermat[2]{\makebox[0pt][l]{$\smash{\underbrace{\phantom{\begin{matrix}#2\end{matrix}}}_{\text{$#1$}}}$}#2}

\journal{Journal of Sound and Vibration}

\begin{document}
	
	\begin{frontmatter}
		
		\title{Applying a Legendre collocation method based on domain decomposition to calculate underwater sound propagation in a horizontally stratified environment}
		
		\author{Houwang Tu}
		\ead{tuhouwang96@163.com}
		\author{Yongxian Wang\corref{cor1}}
		\ead{yxwang@nudt.edu.cn}
		\author{Qiang Lan}
		\author{Wei Liu}
		\author{Wenbin Xiao}
		\author{Shuqing Ma}
		
		\cortext[cor1]{Corresponding author}
		\address{College of Meteorology and Oceanography, National University of Defense Technology, Changsha, China}
		
		\begin{abstract}
			The propagation of sound waves in a horizontally stratified environment, a classic problem in ocean acoustics, has traditionally been calculated using normal modes. Most programs based on the normal mode model are discretized using the finite difference method (FDM). In this paper, a Legendre collocation method (LCM) based on domain decomposition is proposed to solve this problem. A set of collocation points cannot penetrate multiple layers of media, thus necessitating domain decomposition and the use of multiple sets of collocation points. The solution process of this method proceeds entirely in physical space, requiring that the original differential equation be strictly established at the collocation points; thus, a dense matrix eigenvalue system is formed, from which the solution for the horizontal wavenumbers and modes can be directly obtained. Numerical experiments are presented to demonstrate the validity and applicability of this method. A comparison with other methods shows that the LCM proposed in this article is more accurate than the FDM and offers roughly the same accuracy as but a faster calculation speed than other types of spectral methods.
		\end{abstract}
		
		
		\begin{highlights}
			\item Research highlight 1: This paper proposes a new Legendre collocation method based on domain decomposition to calculate acoustic propagation in a horizontally stratified marine environment.
			\item Research highlight 2: The accuracy of the proposed method is greater than that of the classic finite difference method and equivalent to or greater than that of the existing spectral methods.
			\item Research highlight 3: Regarding the run time, the Legendre collocation method is far faster than the Legendre-Galerkin and Chebyshev-Tau spectral methods.
		\end{highlights}
		
		
	\end{frontmatter}
	
	\newpage
	\section{Introduction}
	Sound propagation in a horizontally stratified environment is a subject of enduring interest in underwater acoustics. Pekeris' research article from 1948 first introduced the use of normal modes to solve the layered Pekeris waveguide problem (involving an ocean layer and a sediment layer, with a constant sound speed in each layer) \cite{Pekeris1948}. Subsequently, many computational ocean acoustic models were developed based on normal modes \cite{Finn2011}. One of the most mature normal mode programs is Kraken \cite{Porter2001}, which uses the finite difference method (FDM) to solve the discrete governing equation and can calculate not only the behavior in layered media but also the results for range-dependent situations.
	
	The underwater propagation of sound waves is governed by a set of differential equations. To obtain numerical solutions to partial differential equations, a spectral method is a reliable option that has been widely used in various fields of engineering technology \cite{Zheng1989,Muravskaya2019}. In essence, the spectral method is a variant of the weighted residual method. The variable to be solved is first expanded by selecting a set of special basis functions and obtaining a set of corresponding expanded coefficients. Then, the weighted residuals between the differential equations in the spectral form and the original equation are set equal to 0 to form a set of linear equations in terms of the spectral coefficients of the variable that is sought. Once these linear equations are solved, they are merged into the original variables \cite{Gottlieb1977}. In general, the calculation process of a spectral method is performed in the spectral space \cite{Canuto1988}, where the amount of calculations required are relatively large and the calculation time is relatively long.
	
	In recent years, some scholars have attempted to apply spectral methods to the problem of underwater sound propagation \cite{Adamou2004,Quintanilla2015}. Evans proposed a Legendre-Galerkin technique to solve differential eigenvalue problems with complex and discontinuous coefficients in underwater acoustics \cite{Evans2016}. This method requires that the lower boundary be perfectly free, and the calculation speed is slow. We have presented normal mode and standard parabolic equation models using the Chebyshev-Tau spectral method to process a single layer of a body of water with constant density and no attenuation \cite{Tu2020a}. Subsequently, we extended the method to solve the problem for a layered ocean environment \cite{Tu2020b}. The results showed that the Chebyshev-Tau spectral method offers higher computational accuracy than the FDM, but in terms of run time, the Chebyshev-Tau method is slower than the FDM. Therefore, it would be valuable to develop more efficient computational methods.
	
	Among the available spectral methods, including the Galerkin, Tau, and collocation methods, the collocation method that has been proven to be very fast \cite{Jieshen2011}. This is because in the collocation method, the Dirac-$\delta$ function is used as the weight function when forming a linear system of equations, and the original equation is strictly established at the collocation points but nowhere else \cite{Boyd2001}. The collocation method has the advantages of high precision and high speed and is widely used in various fields of engineering technology \cite{Nuttawit2018,Sinan2020,Wang2019,Khaneh2014}, including acoustics. Chen et al. presented a theoretical formulation based on the collocation method for the eigenanalysis of arbitrarily shaped acoustic cavities \cite{Chen2002}. Li et al. investigated a fast collocation method for acoustic scattering in shallow oceans \cite{Li2008}. Armin et al. analyzed two collocation schemes for the Helmholtz equation with a depth-dependent sound wave velocity, modeling time-harmonic acoustic wave propagation in a three-dimensional inhomogeneous ocean of finite height \cite{Armin2016}. Xu et al. proposed a radial basis function collocation method to solve the fractional Laplacian viscoacoustic wave equation for Earth media with a heterogeneous velocity model and complex geometry \cite{Xu2019}. Wise et al. used Fourier collocation methods to represent arbitrary acoustic source and sensor distributions \cite{Wise2019}. Colbrook et al. analyzed a spectral collocation method for acoustic scattering induced by multiple elastic plates \cite{Colbrook2019}.
	
	The spectral collocation method has been applied in acoustic research, but it is rarely invoked in research on underwater acoustic propagation. Wang et al. developed an application of a Chebyshev collocation method to solve a parabolic equation model of underwater acoustic propagation \cite{Tu2020c}. In 2019, Roberto Sabatini et al. devised a multidomain Chebyshev collocation method for the accurate computation of normal modes in open oceanic and atmospheric waveguides \cite{Sabatini2019}. They set the bottom layer as a semi-infinite layer and then derived a quadratic eigenvalue problem. In this article, we introduce an efficient Legendre collocation method (LCM) that can accommodate the discontinuity of the sound speed, density and attenuation profiles based on normal modes. Considering the discontinuity of these profiles at the interface between the water columns and the bottom sediments, domain decomposition in the depth direction is applied \cite{Min2005,Wu2018}. The structure of this article is as follows. The second section introduces the horizontally stratified ocean environment and the acoustic governing equations. The third section introduces the LCM and the corresponding discretized normal mode model. The fourth section presents test examples and analysis. The last section summarizes the conclusions.
	
	\section{Normal mode model in a layered marine environment}
	\label{section2}
	Considering a cylindrical coordinate system, the acoustic governing equation (Helmholtz equation) of a time-independent harmonic point source is as follows \cite{Finn2011}:
	\begin{equation}
	\label{eq:1}
	\frac{1}{r}\frac{\partial}{\partial r}\left(
	r\frac{\partial p}{\partial r}
	\right) + 
	\rho (z) \frac{\partial}{\partial z}\left(
	\frac{1}{\rho (z)}\frac{\partial p}{\partial z}
	\right) +
	\frac{\omega ^2}{c^2(z)}p=-\frac{\delta(r)\delta(z-z_s)}{2\pi r}
	\end{equation}
	where $p$ is the sound pressure; $\omega=2\pi f$ is the angular frequency of the sound source, where $f$ is the frequency of the source; $r$ is the range, $z$ is the depth; $\rho(z)$ is the density; and $c(z)$ is the speed of sound; $z_s$ is the depth of sound source. Through the separation of variables, the sound pressure can be decomposed as follows:
	\begin{equation}
	\label{eq:2}
	p(r,z)=\psi(z)R(r)
	\end{equation}
	where $R(r)$ may be expressed as a first kind Hankel function (the choice of first kind $H_0^{(1)}(\cdot)$ or second kind $H_0^{(2)}(\cdot)$ is determined by the radiation condition stating that energy should be radiating outward as $r \rightarrow \infty$ \cite{Finn2011}), $\psi (z)$ satisfies the following modal equation:
	\begin{equation}
	\label{eq:3}
	\rho(z)\frac{\mathrm{d}}{\mathrm {d}z}\left(
	\frac{1}{\rho(z)}\frac{\mathrm{d}\psi(z)}{\mathrm {d}z}
	\right) + k^2\psi(z) = k_r^2 \psi(z),\quad k=(1+i\eta\alpha(z))\omega/c(z)
	\end{equation}
	where $k_r^2$ is a constant, with $k_r$ being the horizontal wavenumber; $k$ is the complex wavenumber; the function $\alpha(z)$ describes the attenuation in units of dB$/\lambda$, with $\lambda$ being the wavelength of the sound source; and $\eta=(40\pi \log_{10}{e})^{-1}$ is a constant. The modal equation Eq.~\eqref{eq:3} is a classic Sturm–Liouville eigenvalue problem, the horizontal wavenumbers $k_r$ are effectively eigenvalues for the corresponding modes (eigenfunctions). 
	
	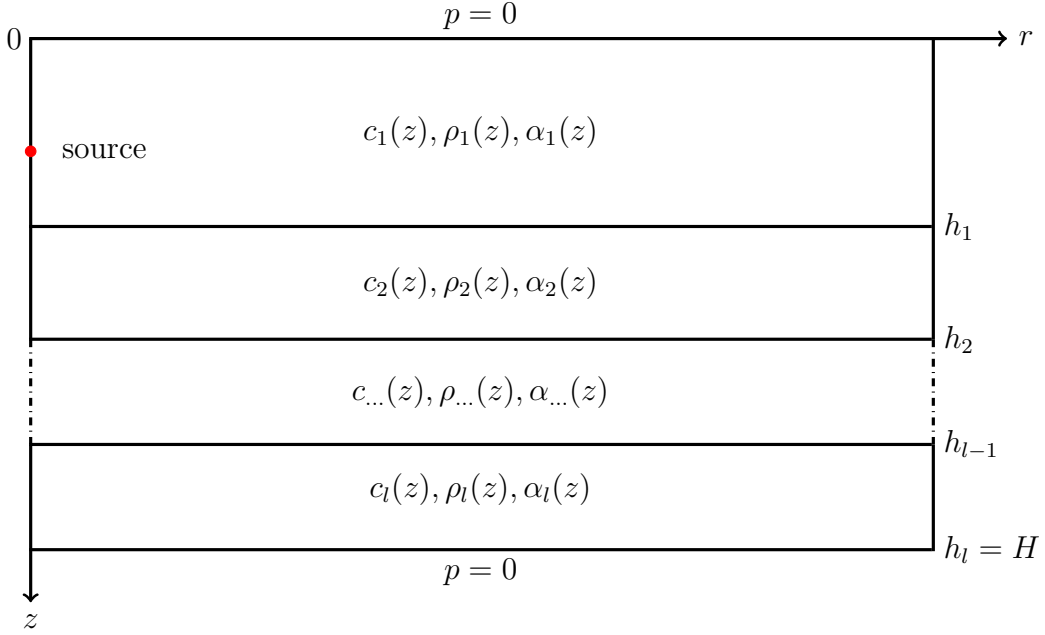
\begin{figure}[htbp]
		\centering
	    \begin{tikzpicture}[node distance=2cm]
    		\node at (1.8,0){$0$};
    		\draw[very thick, ->](2,0)--(15,0) node[right]{$r$};
    		\draw[very thick, ->](2.02,-5.4)--(2.02,-7.5) node[below]{$z$};
    		\draw[very thick](2,-2.5)--(14,-2.5) node[right]{$h_1$};
    		\draw[very thick](2,-4.0)--(14,-4.0) node[right]{$h_2$};
    		\draw[very thick](2,-5.4)--(14,-5.4) node[right]{$h_{l-1}$};
    		\draw[very thick](2,-6.8)--(14,-6.8) node[right]{$h_l=H$};
    		\draw[very thick](2.02,0)--(2.02,-4.0);
    		\filldraw [red] (2.02,-1.5) circle [radius=2pt];
    		\node at (3,-1.5){$\text{source}$};
    		\draw[dash dot, very thick](2.02,-4.0)--(2.02,-5.4);
    		\draw[very thick](14.02,0)--(14.02,-4.0);
    		\draw[dash dot, very thick](14.02,-4.0)--(14.02,-5.4);
    		\draw[very thick](14.02,-5.4)--(14.02,-6.82);
    		\node at (8,-1.25){$c_1(z),\rho_1(z),\alpha_1(z)$};
    		\node at (8,-3.25){$c_2(z),\rho_2(z),\alpha_2(z)$};
    		\node at (8,-4.7){$c_{\dots}(z),\rho_{\dots}(z),\alpha_{\dots}(z)$};
    		\node at (8,-6){$c_l(z),\rho_l(z),\alpha_l(z)$};
    		\node at (8,0.3){$p=0$};
    		\node at (8,-7.1){$p=0$};
	    \end{tikzpicture}
		\caption{Schematic diagram of the horizontally stratified environment.}
		\label{Figure1}
	\end{figure}
	
	The horizontally stratified environment considered herein is a case of cylindrically symmetric two-dimensional sound propagation \cite{Finn2011}. The ocean is divided into $l$ layers of discontinuous water columns or sediments. The environmental parameters are defined separately for the layers, as shown in Fig.~\ref{Figure1}. All the layers satisfy a governing equation of the form given in Eq.~\eqref{eq:1}. 
	
	Boundary conditions are imposed at the limits of the interval $0 \leq z \leq H$, and interface conditions are imposed at $\{h_i\}_{i=1}^l$. The upper boundary condition as follows:
	\begin{equation}
	\label{eq:4}
	p(z=0)=0
	\end{equation}
	The lower boundary condition is:
	\begin{equation}
	\label{eq:5}
	p(z=H)=0
	\end{equation}
	In \cite{Tu2020b}, we also considered the perfectly rigid lower boundary. Sabatini considered the lower boundary condition of a uniform semi-infinite layer in \cite{Sabatini2019}. In fact, if the number of layers is large enough, the perfectly free lower boundary can cope with most situations because the variable number of layers means that any number of absorption layers can be set; and the other two types of boundary conditions will complicate the problem and reduce the efficiency of the algorithm. At the interface, both the sound pressure and the normal particle velocity must be continuous; thus, two constraints are imposed on $p$ at the $l$ interfaces as follows:
	\begin{equation}
	\label{eq:6}
	p(z=h_i^{-})=p(z=h_i^{+}) 
	\end{equation}
	\begin{equation}
	\label{eq:7}
	\frac{1}{\rho(z=h_i^{-})} \frac{\mathrm{d}p(z=h_i^{-} )}{\mathrm {d}z}= \frac{1}{\rho(z=h_i^{+})}\frac{\mathrm{d}p(z=h_i^{+} )}{\mathrm {d}z}
	\end{equation}
	where the superscripts $-$ and $+$ indicate the limits from above and below, respectively.
	
	Supplemented by the boundary and interface conditions in the horizontally stratified environment, Eq.~\eqref{eq:3} has a set of solutions of the form $(k_{r_m},\psi_m)$, $m=1, 2, \dots$, where $\psi_m$ is also called an eigenmode. The eigenmodes of Eq.~\eqref{eq:3} are arbitrary up to a nonzero scaling constant. We shall assume that the modes are normalized such that:
	\begin{equation}
	\label{eq:8}
	\int_{0}^{H} \frac{{[\psi_m(z)}]^2}{\rho(z)}\mathrm {d} z=1,
	\quad m = 1, 2, \dots
	\end{equation}
	Finally, the approximate solution of the 2D Helmholtz equation can be written as follows:
	\begin{equation}
	\label{eq:9}
	p(r,z) \approx \frac{i}{4\rho(z_s)}\sum_{m=1}^{M}\psi_m(z_s)\psi_m(z)H_0^{(1)}(k_{r_m}r)
	\end{equation}
	where $z_s$ is the depth of the sound source, $m$ is the order of mode, $M$ is the number of the modes (Exact sound pressure needs to be obtained by the accumulation of an infinite number of modes. In actual calculations, $M$ modes are usually truncated for approximation.), and $H_0^{(1)}(\cdot)$ is the Hankel function corresponding to $R(r)$ in Eq.~\eqref{eq:2} \cite{Tu2020c}.
	
	\section{Legendre collocation method for the normal modes}
	\subsection{Legendre collocation method}
	Sound propagation in a horizontally stratified ocean environment is solved using the Legendre collocation method (LCM). The collocation method is a spectral method derived from the weighted residual method. In this section, brief descriptions of the fundamental principles of the weighted residual method and the LCM are presented; a more detailed and rigorous derivation can be found elsewhere \cite{Jieshen2011,Canuto2006,Boyd2001}.
	
	To demonstrate the process of solving equations using a spectral method, a one-dimensional differential equation boundary value problem is considered as an example:
	\begin{equation}
	\label{eq:10}
	\begin{split}
	& \mathcal{L}u(x)-f(x)=0,\quad x\in \Omega \backslash \partial \Omega\\
	& \mathcal{B}u(x)-g(x)=0,\quad x\in \partial \Omega
	\end{split}
	\end{equation}
	where $\mathcal{L}$ is a differential operator, $u(x)$ is an unknown function, $\Omega$ is the definite domain, $\mathcal{B}$ is a linear boundary operator, and $\partial \Omega$ is the boundary of $\Omega$. In the method of weighted residuals, a set of linearly independent basis functions is selected to expand the unknown function as follows:
	\begin{equation}
	\label{eq:11}
	u(x)=\sum_{k=0}^{\infty}a_k\phi_k(x)
	\end{equation}
	where $a_k$ are the expansion coefficients and $\phi_k(x)$ are the basis functions. In actual numerical calculations, the number of basis functions used cannot be infinite, so they must be truncated to $(N+1)$ basis functions, as shown in the following formula:
	\begin{equation}
	\label{eq:12}
	u(x)\approx \tilde{u}(x)=\sum_{k=0}^{N}a_k\phi_k(x)
	\end{equation}
	When the expanded $\tilde{u}(x)$ is substituted into Eq.~\eqref{eq:10}, Eq.~\eqref{eq:10} no longer strictly holds, and residuals $R(x)$ of the following form arise:
	\begin{equation}
	\label{eq:13}
	R(x)=\mathcal{L}\tilde{u}(x)-f(x)
	\end{equation}
	A constraint is applied to the expansion coefficients by setting the weighted residuals to 0 as follows:
	\begin{equation}
	\label{eq:14}
	\int_{\Omega} R(x)w_j(x) \mathrm {d}x=0,\quad j=0,1,2,\dots,N
	\end{equation}
	where $w_j(x)$ are the weight functions. Then, the expansion coefficients can be obtained via Eq.~\eqref{eq:14}. Therefore, the choices of the basis functions $\{\phi_k(x)\}_{k=0}^N$ and the weight functions $\{w_j(x)\}_{j=0}^N$ are very important. The collocation method is a spectral method in which the $\delta$ function is used as the form of the weight functions:
	\begin{equation}
	\label{eq:15}
	\int_{\Omega} R(x) \delta(x-x_j) \mathrm {d}x
	= {R(x_j)
		= \mathcal{L}\tilde{u}(x_j)-f(x_j)} =0,
	\quad j=0,1,2,\dots,N
	\end{equation}
	where $x_j$ is the $j$-th point among $(N+1)$ discrete grid points in $\Omega$. As seen from the above formula, the collocation method essentially requires the weighted residuals to be 0 only at the specified discrete points, and the equations established to solve for unknown functions are expressed in the physical space instead of the spectral space. In other words, the collocation method requires that the original equation be strictly established only at discrete grid points $x_j$ and not elsewhere in $\Omega$. The basis functions of the LCM are the Legendre orthogonal polynomials \cite{Boyd2001}, which are defined on the interval $x \in [-1, 1]$ as follows:
	\begin{equation}
	\label{eq:16}
	\begin{split}
	&P_0(x)=1,\quad P_1(x)=x\\
	(n+1)P_{n+1}(x)&=(2n+1)xP_n(x)-nP_{n-1}(x),\quad n\ge 1
	\end{split}
	\end{equation}
	where $x \in [-1, 1]$ can be simply scaled to any bounded interval $\Omega$. In LCM, the $(N+1)$ discrete grid points $x_j$ on the interval $[-1,1]$ can be used as the nonequidistant Legendre-Gauss-Lobatto (LGL) collocation points, $x_0=-1$, $x_N=1$, and $\{x_j\}_{j=1}^{N-1}$ are the zeros of $P_N^{'}(x)$ \cite{Gottlieb1977}. In the solution process of the LCM, all functions are represented by their values at the $(N+1)$ LGL points. Accordingly, each function in the solution obtained via the LCM can be expressed as an $(N+1)$-dimensional column vector as follows:
	\begin{equation}
	\label{eq:17}
	\begin{split}
	\mathbf{u}&=[u_0, u_1, u_2,\dots, u_N]^{\mathrm{T}},\quad u_j=u(x_j),\\
	\mathbf{f}&=[f_0, f_1, f_2,\dots, f_N]^{\mathrm{T}},\quad f_j=f(x_j),\quad j=0,1,\dots,N
	\end{split}
	\end{equation}
	
	To apply the LCM to the differential equations given in Eq.~\eqref{eq:10}, the discretization of the operators $\mathcal{L}$ and $\mathcal{B}$ are necessary. The operators may contain derivatives, multiplications, and/or integrations.
	\begin{enumerate}
		\item
		Derivatives
		
		The derivative $u'(x)$ of the function $u(x)$ is sometimes included in the $\mathcal{L}$ operator; we express the relation between the two as follows:
		\begin{equation}
		\label{eq:18}
		\mathbf{u}'=\mathbf{D}_N\mathbf{u}
		\end{equation}
		where $\mathbf{u}'=[u'_0,u'_1, u'_2,\dots, u'_N]^{\mathrm{T}}$ represents the values of the derivative function $u'(x)$. The matrix $\mathbf{D}_N$, called the Legendre collocation differential matrix, describes the relation between the original function $u(x)$ and its first derivative in the LCM. For the derivation of the matrix $\mathbf{D}_N$, the interested reader refer to Eq.~(2.3.25) to Eq.~(2.3.28) in Monograph \cite{Canuto2006}. All the elements in the matrix $\mathbf{D}_N$ can be calculated as follows:
		\begin{equation}
		\label{eq:19}
		(D_N)_{kj}=
		\begin{cases}
		\frac{P_N(x_k)}{P_N(x_j)}\frac{1}{x_k-x_j},& k\ne j;\\
		-\frac{N(N+1)}{4}, & k = j = 0; \\
		\frac{N(N+1)}{4}, & k = j = N; \\
		0, & \text{otherwise}
		\end{cases}
		\end{equation}
		
		\item
		Products
		
		Let the product of the functions $v(x)$ and $u(x)$ be $y(x)=v(x)u(x)$. Then, $y(x)$ can be processed as follows in the LCM:
		\begin{equation}
		\label{eq:20}
		\mathbf{y}=\mathbf{C}_v\mathbf{u}
		\end{equation}
		where $\mathbf{C}_v$ is an $(N+1)\times(N+1)$ diagonal matrix and $(C_v)_{ii}=v_i=v(x_i)$, $i=0,1,\dots,N$.
		\item
		Integrals
		
		The integral $\int_{-1}^{1}u(x) \mathrm {d} x$ can be obtained by means of the Gauss-Legendre-Lobatto quadrature \cite{Jieshen2011}. The weights of the Gauss-Legendre-Lobatto quadrature are as follows:
		\begin{equation}
		\label{eq:21}
		w_j=\frac{2}{N(N+1)}\frac{1}{[P_N(x_j)]^2},\quad 1\le j \le N
		\end{equation}
		
	\end{enumerate}
	
	In this way, a differential equation can be discretized into a system of linear equations, and the problem can be solved. In actual numerical calculations, the unknown function $u(x)$ is not expanded with Legendre polynomials; instead, the original equation is constrained to be strictly true at the LGL points, and this constraint is ultimately used to solve the linear equations to directly obtain the function values $u(x_j)$ at the LGL points $x_j$. Since the LGL points are not equidistant, the function values at the remaining points on the interval $x$ can be obtained through interpolation once $u(x_j)$ are obtained.
	
	A systematic theoretical study of the convergence of the Legendre polynomials and the collocation method was presented by Boyd and Shen \cite{Boyd2001,Jieshen2011}. The asymptotic rate of convergence in the LCM has been previously reported; more details are provided by Canuto \cite{Canuto1988,Canuto2006}.
	
	\subsection{Discrete normal mode model using the Legendre collocation method}
	We first consider a single-layer environment model without discontinuous layers. The domain of the problem solved by LCM is usually in the interval $[-1,1]$. To apply the LCM in the normal mode model, we use
	\[
        x=\frac{2z}{b-a}-\frac{b+a}{b-a},\quad \frac{\mathrm {d}x}{\mathrm {d}z}=\frac{2}{b-a}
	\]
	to scale the domain $z\in[a,b]\mapsto x\in[-1,1]$. Eq.~\eqref{eq:3} has the following forms: 
	\begin{equation}
	\label{eq:22}
	\frac{4}{(b-a)^2}\rho(x)\frac{\mathrm{d}}{\mathrm {d}x}\left(
	\frac{1}{\rho(x)}\frac{\mathrm{d}\psi(x)}{\mathrm {d}x}
	\right) + k^2\psi(x) = k_r^2 \psi(x)
	\end{equation}
	We introduce the discretization of the first term of Eq.~\eqref{eq:24} in the LCM. Let $v(x)$, $g(x)$, and $s(x)$ be
	\begin{equation}
	\label{eq:23}
	\begin{split}
	&v(x)= \frac{1}{\rho(x)}, \quad g(x)=v(x)\frac{\mathrm{d}\psi(x)}{\mathrm {d}x},\\
	&s(x)=\rho(x) g'(x) = \rho(x) \frac{\mathrm{d}}{\mathrm {d}x}\left[\frac{1}{\rho(x)}\frac{\mathrm{d}\psi(x)}{\mathrm {d}x}\right]
	\end{split}
	\end{equation}
	After $(N+1)$-order truncation, we obtain the discrete first term on the left side of Eq.~\eqref{eq:22} as follows:
	\begin{equation}
	\label{eq:24}
	\mathbf{g}=\mathbf{C}_v(\mathbf{D}_N\bm{\psi}),\quad
	\mathbf{s}=\mathbf{C}_{\rho}(\mathbf{D}_N\mathbf{g})
	=\mathbf{C}_{\rho}(\mathbf{D}_N(\mathbf{C}_v(\mathbf{D}_N\bm{\psi}))
	\end{equation}
	Therefore, the discrete form of Eq.~\eqref{eq:24} in the LCM is:
	\begin{equation}
	\label{eq:25}
	\left[\frac{4}{(b-a)^2}\mathbf{C}_{\rho}\mathbf{D}_N\mathbf{C}_{v}\mathbf{D}_N+\mathbf{C}_{k^2}\right]\bm{\psi}=k_r^2\bm{\psi}
	\end{equation}
	where $\mathbf{C}_{\rho}$, $\mathbf{C}_{v}$ and $\mathbf{C}_{k^2}$ have a meaning similar to that of $\mathbf{C}_v$ in Eq.~\eqref{eq:20} and $\mathbf{D}_N$ has a meaning similar to that of $\mathbf{D}_N$ in Eq.~\eqref{eq:19}.
	
	As shown in Fig.~\ref{Figure1}, a single set of basis functions cannot span many layers since the outcome would not have the required derivative discontinuity at the interfaces $\{h_i\}_{i=1}^l$. Thus, it is advisable to use the domain decomposition method of Min \cite{Min2005} in Eq.~\eqref{eq:3} of this article. We split the domain interval into $l$ subintervals. For every splitting event, the associated discontinuous point is the endpoint of one subinterval. We then take the $(N_i+1)$ LGL points $z_{i,j}$ on $z_i \in [h_{i-1},h_i]$, where $i=1,2,\cdots,l; j=0,1,\cdots,N_i$. For convenience of description, we adopt the following notations analogous to Eq.~\eqref{eq:17}:
	\begin{equation}
	\label{eq:26}
	\begin{split}
	&k_i=k(z_i),\quad \rho_{i}=\rho(z_i), \quad \rho_{i,j}=\rho(z_{i,j})\\ &\gamma_{i,1}=\frac{2}{h_i-h_{i-1}}\frac{1}{\rho_{i,N_i}},\quad \gamma_{i,2}=\frac{2}{h_i-h_{i-1}}\frac{1}{\rho_{i,0}}, \quad h_0=0\\
	&\bm{\psi}_i=[\psi_{i,0},\psi_{i,1},\psi_{i,2},\dots,\psi_{i,N_i}]^\mathrm{T},\quad  \psi_{i,j}=\psi(z_{i,j})\\
	&\bm{\psi}=[
	\psi_{1,0},\psi_{1,1},\cdots,\psi_{1,N_1},
	\psi_{2,0},\psi_{2,1},\cdots,\psi_{2,N_2},
	\cdots,
	\psi_{l,0},\psi_{l,1},\cdots,\psi_{l,N_l}
	]^{\mathrm{T}}
	\end{split}
	\end{equation}
	As a result, the matrix form of Eq.~\eqref{eq:3} in the $i$-th layer is rewritten as follows:
	\begin{equation}
	\label{eq:27}
	\left[\frac{4}{(h_i-h_{i-1})^2}\mathbf{C}_{\rho_i}\mathbf{D}_{N_i}\mathbf{C}_{1/\rho_i}\mathbf{D}_{N_i}+\mathbf{C}_{k_i^2}\right]\bm{\psi}_i=k_r^2\bm{\psi}_i
	\end{equation}
	
	Let the linear matrix operators on the left-hand side of Eq.~\eqref{eq:27} be $\mathbf{W}_i$. Considering that the interface conditions are related to both the upper layer and the lower layer, we need to solve $l$ Eqs.~\eqref{eq:27} together, as expressed by the following equation:
	\begin{equation}
	\label{eq:28}
	\left[\begin{array}{cccc}
	\mathbf{W}_1&\mathbf{0}&\mathbf{0}&\mathbf{0}\\
	\mathbf{0}&\mathbf{W}_2&\mathbf{0}&\mathbf{0}\\
	\mathbf{0}&\mathbf{0}&\ddots&\mathbf{0}\\
	\mathbf{0}&\mathbf{0}&\mathbf{0}&\mathbf{W}_l
	\end{array}\right]
	\left[\begin{array}{c}
	\bm{\psi}_1\\
	\bm{\psi}_2\\
	\vdots\\
	\bm{\psi}_l\\
	\end{array}
	\right]=k_r^2\left[\begin{array}{c}
	\bm{\psi}_1\\
	\bm{\psi}_2\\
	\vdots\\
	\bm{\psi}_l\\
	\end{array}
	\right]
	\end{equation}
	With respect to the boundary and interface conditions in the LCM, let the first row of the matrix $\mathbf{D}_{N_i}$ be the row vector $\mathbf{s}_i$, and let the last row of $\mathbf{D}_{N_i}$ be the row vector $\mathbf{q}_i$. Then, the boundary and interface conditions in Eqs.~\eqref{eq:4} to \eqref{eq:7} can be modified:
	\begin{equation}
	\label{eq:29}
	\psi_{1,0}=0
	\end{equation}
	\begin{equation}
	\label{eq:30}
	\psi_{l,N_l}=0
	\end{equation}
	\begin{equation}
	\label{eq:31}
	\psi_{h_i,N_i}-\psi_{h_{i+1},0}=0
	\end{equation}
	\begin{equation}
	\label{eq:32}
	\gamma_{i,1}\mathbf{q}_i\bm{\psi}_i-\gamma_{i+1,2}\mathbf{s}_{i+1}\bm{\psi}_{i+1}=0
	\end{equation}
	
	Let $N = \sum_{i=1}^lN_i$ and let the square matrix of $(N+l)$-order on the left-hand side of Eq.~\eqref{eq:28} be denoted by $\mathbf{L}$. The first row of the matrix $\mathbf{L}$ is replaced with the $(N+l)$-dimensional row vector $[1,0,0,\dots,0]$, and the last row of $\mathbf{L}$ is replaced with the $(N+l)$-dimensional row vector $[0,0,0,\dots,1]$. Similarly, Eqs.~\eqref{eq:31} and \eqref{eq:32} are also used to replace the corresponding rows of the $\mathbf{W}_i$ and $\mathbf{W}_{i+1}$ matrices. The corresponding position of $\bm{\psi}$ on the right-hand side of Eq.~\eqref{eq:28} should also be replaced with 0; the specific positions are $\psi_{i,0}$ and $\psi_{i,N_i}$. Let the modified Eq.~\eqref{eq:28} be expressed as:
	\begin{equation}
	\label{eq:33}
	\widetilde{\mathbf{L}}\bm{\psi}= k_r^2\mathbf{b}
	\end{equation}
	where $\mathbf{b}=[0,\psi_{1,1},\cdots,\psi_{1,N_1-1},0,\cdots,0,\psi_{l,1},\cdots,\psi_{l,N_l-1},0]^{\mathrm{T}}$. To facilitate the solution of Eq.~\eqref{eq:33}, we rearrange the square matrix $\widetilde{\mathbf{L}}$ of the $(N+l)$-th order on the left-hand side of Eq.~\eqref{eq:33} by shifting the $2l$ replaced rows to the last $2l$ rows. Then, the matrix $\overline{\mathbf{L}}$ obtained in the previous step is divided into blocks as follows:
	\begin{equation}
	\label{eq:34}
	\overline{\mathbf{L}}
	=\left[\begin{array}{c|c}
	\mathbf{L}_{11}  & \mathbf{L}_{12} \\
	\hline
	\undermat{N-l}{\mathbf{L}_{21}} &
	\undermat{2l}{\mathbf{L}_{22}}
	\end{array}\right]
	\begin{array}{@{} l @{}}
	\left.
	\vphantom{
		\begin{array}{c}
		\mathbf{L}_{11}
		\end{array}
	}
	\right\} N-l \\
	\left.\vphantom{
		\begin{array}{c}
		\mathbf{L}_{11}
		\end{array}
	}
	\right\} 2l
	\end{array}
	\end{equation}
	where $\mathbf{L}_{11}$ is a square matrix of order $(N-l)$ and $\mathbf{L}_{22}$ is a square matrix of order $2l$. The dimensions of $\mathbf{L}_{12}$ are $(N-l)\times 2l$, and the dimensions of $\mathbf{L}_{21}$ are $2l\times (N-l)$. In accordance with the shape of $\overline{\mathbf{L}}$, Eq.~\eqref{eq:34} can be written as the following block matrix equation:
	\begin{equation} 
	\label{eq:35}
	\overline{\mathbf{L}}\bm{\Psi}
	=k_r^2 \overline{\mathbf{b}}
	\quad \text{or} \quad
	\left[\begin{array}{cc}
	\mathbf{L}_{11}&\mathbf{L}_{12}\\
	\mathbf{L}_{21}&\mathbf{L}_{22}\\
	\end{array}
	\right]\left[
	\begin{array}{c}
	\bm{\Psi}_1\\
	\bm{\Psi}_2
	\end{array}
	\right]=k_r^2\left[
	\begin{array}{c}
	\bm{\Psi}_1\\
	\mathbf{0}
	\end{array}\right]
	\end{equation}
	where 
	\[\bm{\Psi}_1=[
	\psi_{1,1},\cdots,\psi_{1,N_1-1},
	\cdots,
	\psi_{l,1},\cdots,\psi_{l,N_l-1}
	]^{\mathrm{T}}
	\]
	and 
	\[\bm{\Psi}_2=[
	\psi_{1,0},\psi_{1,N},
	\cdots,
	\psi_{l,0},\psi_{l,N}
	]^{\mathrm{T}}.
	\]
	Eq.~\eqref{eq:35} can be decomposed into the following equations:
	\begin{equation}
	\label{eq:36}
	\bm{\Psi}_2=-\mathbf{L}_{22}^{-1}\mathbf{L}_{21}\bm{\Psi}_1,\quad(\mathbf{L}_{11}-\mathbf{L}_{12}\mathbf{L}_{22}^{-1}\mathbf{L}_{21})\bm{\Psi}_1=k_r^2\bm{\Psi}_1
	\end{equation}
	Finally, we solve the eigenvalue/eigenvector problem expressed in Eq.~\eqref{eq:36}. By reordering the eigenvectors to their original order and stacking $\bm{\Psi}_1$ and $\bm{\Psi}_2$ into an $(N+1)$-dimensional column vector $\bm{\psi}=[\psi_{1,0},\cdots,\psi_{1,N_1};\cdots;\psi_{l,0},\cdots,\psi_{l,N_l}]^{\mathrm{T}}$, a one-to-one matching of $(N-l)$ eigenvalue/eigenvector pairs is obtained. Then, we use Eq.~\eqref{eq:8} to normalize $\bm{\psi}$; finally, we obtain a set of modes $(k_r,\bm{\psi})$. The modal functions $\bm{\psi}$ thus obtained are established at the nonequidistant LGL points, and approximations of the solution at any other points can be easily interpolated from the obtained modes; for example, the modal function values at the depth of the sound source, $\psi(z_s)$, can be obtained via interpolation to synthesize the sound field. Then, subject to the limitation of the phase speed, we select an appropriate number of modes and synthesize the sound pressure field in accordance with Eq.~\eqref{eq:9}.
	
	\section{Test examples and analysis of results}
	
	To verify the validity of the LCM in solving the normal mode model and to summarize the characteristics of this method, the following tests and analyses were performed based on the 4 representative examples described in Table \ref{tab1}. Example 1 is a simple ideal fluid waveguide, which has an analytical solution and can be used to accurately compare the accuracy of the method proposed in this article with other methods. Example 2 is a layered deep-sea environment, which can be used to verify the correctness of LCM in a deep-sea environment where density varies with depth. Example 3 has a nonsmooth measured sound speed profile, which can be used to demonstrate the practicality of LCM. Example 4 simulates a Pekeris waveguide. This example is used to show that LCM can approximately simulate sound propagation in a semi-infinite space after adding a perfectly absorbing layer.
	
	For comparison, we consider the widely used Kraken program (including kraken.exe and field.exe) based on finite difference discretization, the rimLG program based on the Legendre-Galerkin spectral method \cite{Evans2016}, and the normal modes program based on the Chebyshev-Tau spectral method (NM-CT) \cite{Tu2020c,Tu2021code}. To facilitate the description, our normal mode model program based on the LCM developed in this paper is abbreviated as MultiLC (Multilayer Legendre collocation method). The programs rimLG, NM-CT and MultiLC are all based on spectral methods. 
	
\begin{figure}[htbp]
		\centering
		\subfigure[]{\label{Figure2a}\includegraphics[width=3.3cm]{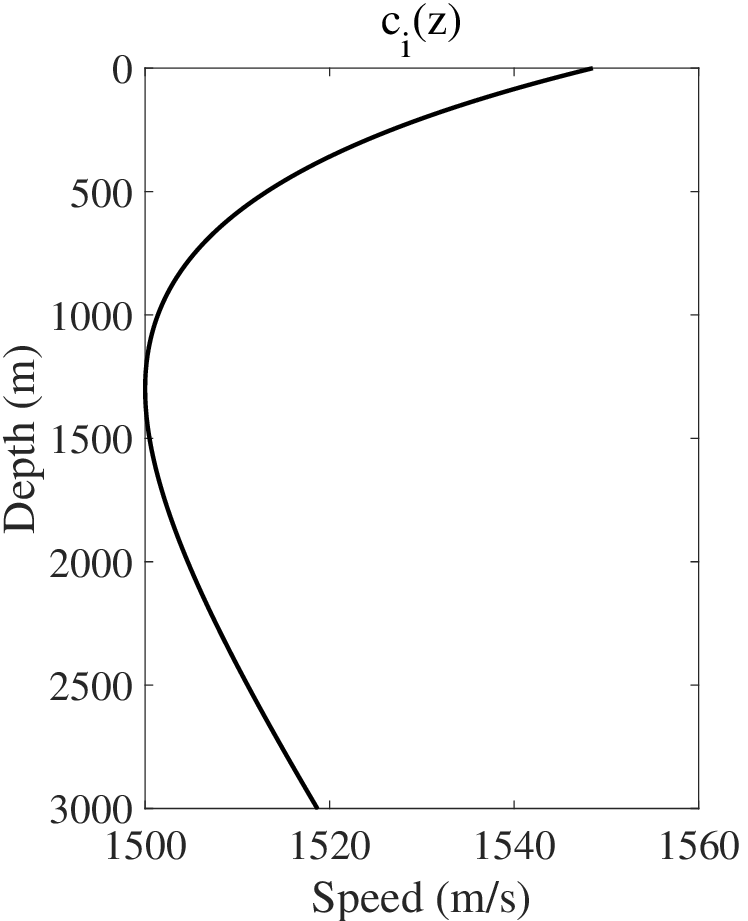}}
		\subfigure[]{\label{Figure2b}\includegraphics[width=3.3cm]{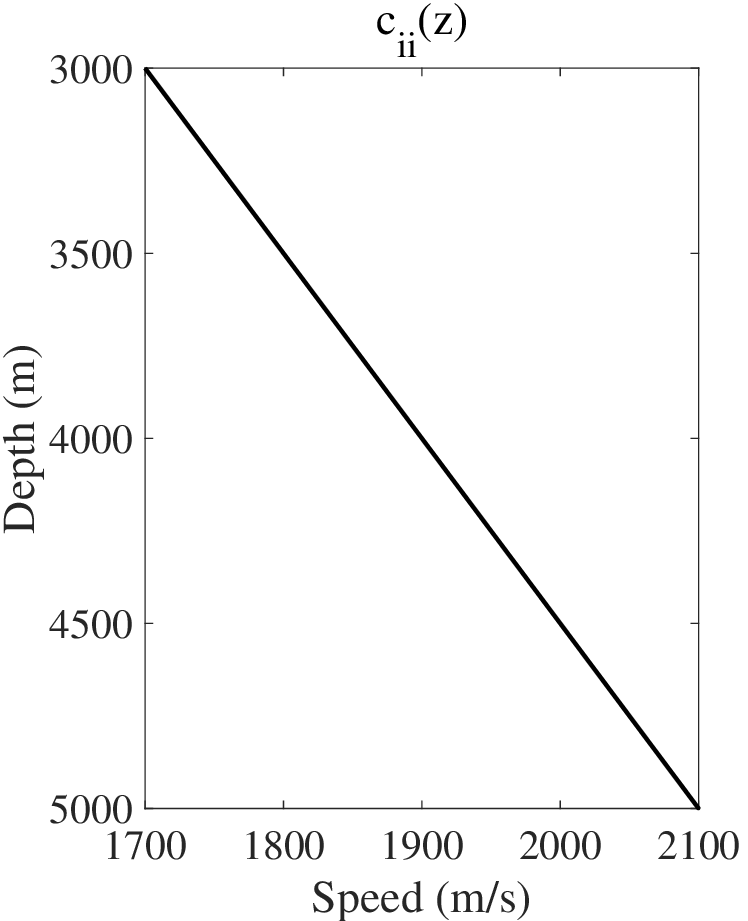}}
		\subfigure[]{\label{Figure2c}\includegraphics[width=3.3cm]{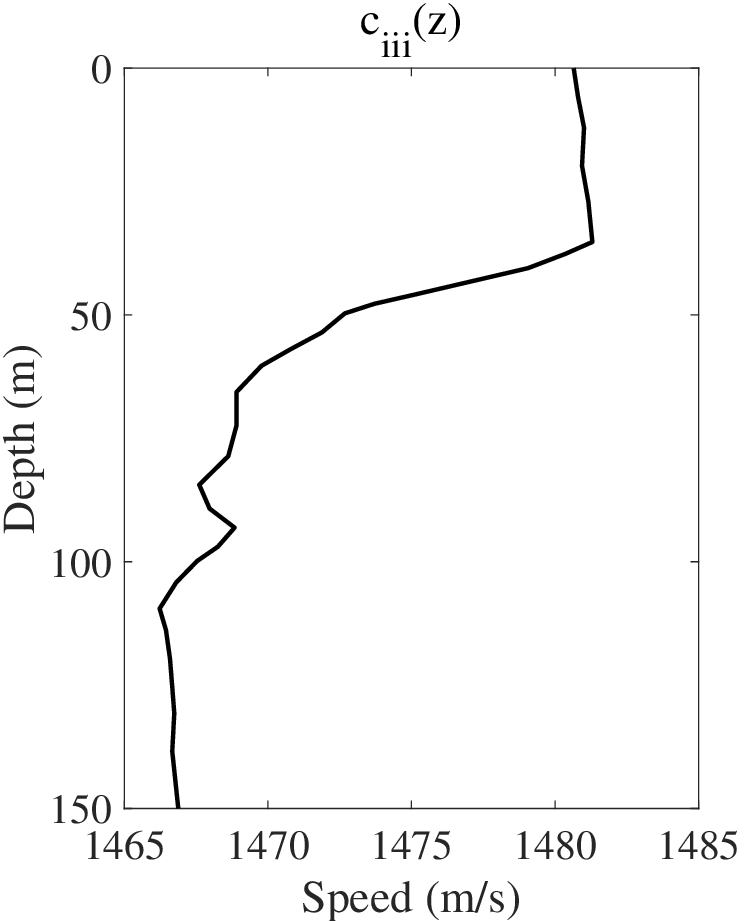}}
		\subfigure[]{\label{Figure2d}\includegraphics[width=3.1cm]{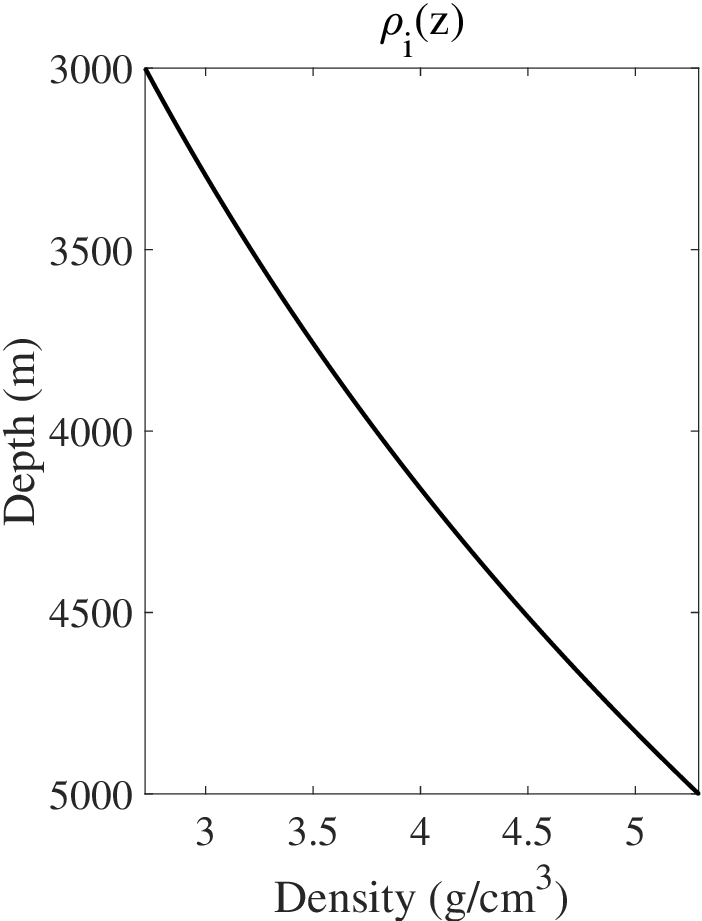}}
		\caption{Acoustic profiles used in the examples.}
	\end{figure}

	The various acoustic profiles used for the numerical examples are as follows:
	\begin{enumerate}
		\item
		The deep water Munk sound speed profile \cite{Finn2011} shown in Fig.~\ref{Figure2a} is given by:
		\begin{equation}
		\begin{split}
		c_1(z)&=1500.0[1.0+\varepsilon(\tilde{z}-1+\exp{(-\tilde{z})})]\\
		\varepsilon&=0.0073,\quad \tilde{z}=(z-1300)/650
		\end{split}
		\end{equation}
		\item
		The linear sound speed profile in the bottom layer shown in Fig.~\ref{Figure2b} is given by:
		\begin{equation}
		c_2(z)=0.2z+1100
		\end{equation}
		\item
		The measured sound speed profile $c_3(z)$ from the Barents Sea reported in \cite{Chowdhury2018} is shown in Fig.~\ref{Figure2c}.
		\item
		The exponentially increasing density profile in the bottom layer shown in Fig.~\ref{Figure2d} is given by:
		\begin{equation}
		\rho_1(z)=\exp \left(\frac{z}{3000}\right)
		\end{equation}
	\end{enumerate}
	
	\begin{table}[htbp]
		\renewcommand\arraystretch{1.5}
		\scriptsize
		\caption{\label{tab1} List of examples.}
		\centering
		\begin{tabular}{ccccccccccccc}
			\hline
			\multirow{2}{*}{No.} &
			${\rho_1}$&
			${c_1}$& 
			${\alpha_1}$& 
			${h_1}$& 
			${\rho_2}$& 
			${c_2}$& 
			${\alpha_2}$ & 
			${h_2}$& 
			${\rho_3}$& 
			${c_3}$& 
			${\alpha_3}$ & 
			${H}$ 
			\\
			& 
			\scriptsize{g/cm$^3$}        &
			\scriptsize{m/s}             &
			\scriptsize{dB/$\lambda$}    &
			\scriptsize{m}               &
			\scriptsize{g/cm$^3$}        &
			\scriptsize{m/s}             &
			\scriptsize{dB/$\lambda$}    &
			\scriptsize{m}               &
			\scriptsize{g/cm$^3$}        &
			\scriptsize{m/s}             &
			\scriptsize{dB/$\lambda$}    &
			\scriptsize{m}          
			\\
			\hline
			{1}& 
			{1.0} &
			{1500} &
			{0.0} &
			{50}&
			{1.0}&
			{1500}& 
			{0.0}&&&&&
			{100}
			\\
			{2}& 
			{1.0} &
			{$c_{\text{i}}$}& 
			{0.0}&
			{3000}&
			{$\rho_{\text{i}}$}&
			{$c_{\text{ii}}$}&
			{0.0}&&&&&
			{5000}
			\\
			{3}& 
			{1.0} &
			{$c_{\text{iii}}$}&   
			{0.0}&
			{150}&
			{2.0}&
			{1800}&
			{0.5}&&&&&
			{300}
			\\
			{4}& 
			{1.0} &
			{1500}&   
			{0.0}&
			{100}&
			{1.5}&
			{2000}&
			{0.5}&
			{200}&
			{1.5}&
			{2000}&
			{5.0}&
			{300}
			\\
			\hline
		\end{tabular}
	\end{table}
	
	To describe the sound field results, the transmission loss (TL) of the sound pressure is defined as $\text{TL}=-20\log_{10}(|p|/|p_0|)$. The TL is expressed in units of decibels (dB), and $p_0$ is the sound pressure at a distance of 1 m from the sound source. In practice, the TL is usually used in actual sound field displays \cite{Finn2011}.
	
	\subsection{Single-layer waveguide}
\begin{figure}[htbp]
		\centering
		\includegraphics[width=12cm]{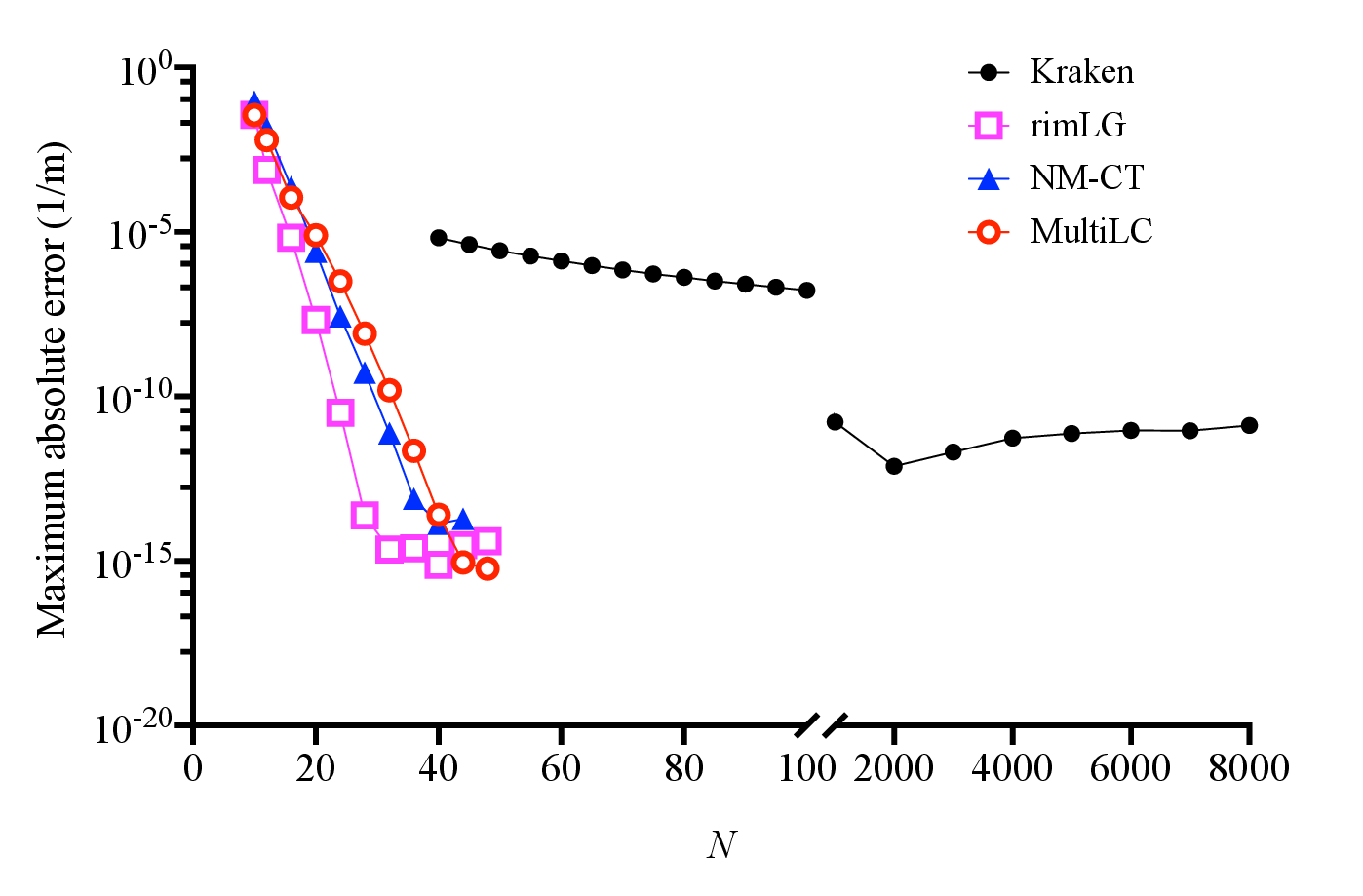}
		\caption{Trends of the maximum absolute errors of the horizontal wavenumbers calculated by the tested programs with an increasing truncation order $N$ at $f$=50 Hz for Example 1.}
		\label{Figure3}
	\end{figure}

	Example 1 in Table \ref{tab1} is a single-layer waveguide problem. The density of the seawater is taken to be constant at 1 $\text{g}/\text{cm}^3$, and there is no attenuation in the water. The LCM proposed in this paper is also suitable for this type of single-layer ocean model; we need only to set the acoustic parameters in the two layers as one continuous layer in the environmental file. According to the wavenumber integration method, the exact analytical solution for Example 1 is:
	\begin{equation}
	\begin{split}
	p(r,z)=\frac{2\pi i}{H}\sum_{m=1}^{\infty}\sin{(k_{z_m}z_s)}\sin{(k_{z_m}z)}H_0^{(1)}(k_{r_m}r),\\
	k_{z_m}=\frac{m\pi}{H},\quad k_{r_m}=\sqrt{k_0^2-k_{z_m}^2}, \quad m=1,2,3,\cdots
	\end{split}
	\end{equation}
	When calculating the numerical sound field, a finite number of $m$ is generally truncated. Table \ref{tab2} shows a few horizontal wavenumbers for Example 1 calculated from the analytical solution and by Kraken, rimLG, NM-CT and MultiLC for two frequencies: $f$ = 20 Hz and 50 Hz. Kraken uses 100 discrete grid points in the vertical direction, whereas rimLG, NM-CT and MultiLC use spectral truncation at orders of $N_i$ = 20 at 20 Hz and $N_i$ = 40 at 50 Hz. The wavenumbers calculated by all 4 programs are very similar to the values from the analytical solution. However, under the current configuration, Kraken results are not as accurate as the results of the other 3 programs, which are essentially identical to the analytical solution, with no error within ten decimal places. Fig.~\ref{Figure3} shows the trends of the maximum absolute errors of the horizontal wavenumbers calculated by Kraken, rimLG, NM-CT and MultiLC as the truncation order $N$ increases at 50 Hz. When the number of discrete points is too few, the Kraken program terminates; the maximum absolute error of the wavenumbers calculated by Kraken decreases approximately quasilinearly as $N$ increases to 100. The maximum absolute errors of the wavenumbers calculated by the other 3 programs rapidly decrease and converge to an extremely low level as $N$ increases to 50. These findings are consistent with the mathematically derived conclusion that the error of a spectral method exhibits exponential convergence. When the truncation order is increased to a sufficient level, the calculation results of the programs become stable; therefore, analyzing the corresponding errors is helpful for determining which method has the highest accuracy. As seen from the figure, MultiLC and rimLG achieve the smallest errors, on the order of $10^{-15}$, whereas NM-CT has a slightly larger error, on the order of $10^{-14}$. Compared to the FDM-based Kraken program, all 3 of these programs represent high-precision methods.
	
	\begin{table}[htbp]
		\renewcommand\arraystretch{1.5}
		\scriptsize
		\caption{\label{tab2}Comparison of $k_{r_m}$ for Example 1.}
		\centering
		\begin{tabular}{ccccccc}
			\hline
			{$f$ (Hz)}   & 
			{$m$}        &
			{Analytical} & 
			{Kraken}	 & 
			{rimLG}      &
			{NM-CT}      &
			{MultiLC}\\
			\hline
			20
			& 1     & 0.07766 22489  & 0.07766 224\textcolor{red}{90} &0.0776622489 &0.07766 22489   &0.07766 22489\\
			& 2     & 0.05541 24859  & 0.05541 248\textcolor{red}{63} &0.0554124859 &0.05541 24859   &0.05541 24859\\
			\hline
			50
			& 1    & 0.20706 99109   & 0.20706 99109 &0.20706 99109 &0.20706 99109  &0.20706 99109\\
			& 2    & 0.19979 25591   & 0.19979 2559\textcolor{red}{2} &0.19979 25591 &0.19979 25591  &0.19979 25591\\
			& 3    & 0.18703 54632   & 0.18703 546\textcolor{red}{45} &0.18703 54632 &0.18703 54632  &0.18703 54632\\
			& 4    & 0.16755 16082   & 0.16755 16\textcolor{red}{163} &0.16755 16082 &0.16755 16082  &0.16755 16082\\
			& 5    & 0.13853 12147   & 0.13853 12\textcolor{red}{523} &0.13853 12147 &0.13853 12147  &0.13853 12147\\
			& 6    & 0.09129 25660   & 0.09129 2\textcolor{red}{7365} &0.09129 25660 &0.09129 25660  &0.09129 25660\\        
			\hline
		\end{tabular}
	\end{table}
	
	To evaluate the speed of the method proposed in this article, Table \ref{tab3} shows the run times for Example 1 at the two frequencies. Kraken is written in FORTRAN; rimLG and NM-CT are written in MATLAB. For the sake of comparison, the authors have developed two versions of MultiLC programs using FORTRAN and MATLAB. The test platform was a Dell XPS 8930 desktop computer equipped with an Intel i7-8700K CPU and 16 GB of memory. The FORTRAN compiler used in the tests was gfortran 7.5.0, and the MATLAB version was 2019a. The results listed in the table are the averages calculated for ten runs after discarding the maximum and minimum values.
	
	\begin{table}[htbp]
		\renewcommand\arraystretch{1.5}
		\scriptsize
		\caption{\label{tab3}Comparison of the run times for Example 1 (unit: s).}
		\centering
		\begin{tabular}{cccccc}
			\hline
			$f$ (Hz) &Kraken &MultiLC &MultiLC (MATLAB) &rimLG (MATLAB) &NM-CT (MATLAB)  \\ 
			\hline
			20  &0.319  &0.036 &0.077 & 7.531 &0.121   \\
			50  &0.379  &0.046 &0.079 & 7.914 &0.183    \\
			\hline
		\end{tabular}
	\end{table}
	
	Based on the run times of the programs for Example 1, rimLG is the slowest, NM-CT is faster than rimLG, and MultiLC is the fastest. This implies that when the results of roughly the same accuracy are obtained, the amount of calculation required in MultiLC is less than that in rimLG or NM-CT, and even less than that of Kraken based on FDM.
	
	\subsection{Layered waveguides}
	Examples 2 and 3 in Table \ref{tab1} are two-layer waveguide problems with bottom attenuation. When there is an attenuating bottom sediment layer, the horizontal wavenumbers are complex. Kraken uses an alternative approach to complex eigenvalue searches in which the real eigenvalue is calculated first, and then an approximation of the imaginary part is obtained using perturbation theory (\cite{Porter2001}, p35, Sect. 2.7.5), which introduces slight inaccuracies in the computed wavenumbers. However, KrakenC is a version of Kraken in which the wavenumbers are found in the complex plane, allowing complex eigenvalues to be computed exactly. Therefore, we use KrakenC instead of Kraken for the examples with attenuation.
	
	Example 2 is a complex two-layer waveguide in which the sound speed and density are not constant.
	
\begin{figure}[htbp]
		\centering
		\includegraphics[width=\linewidth] {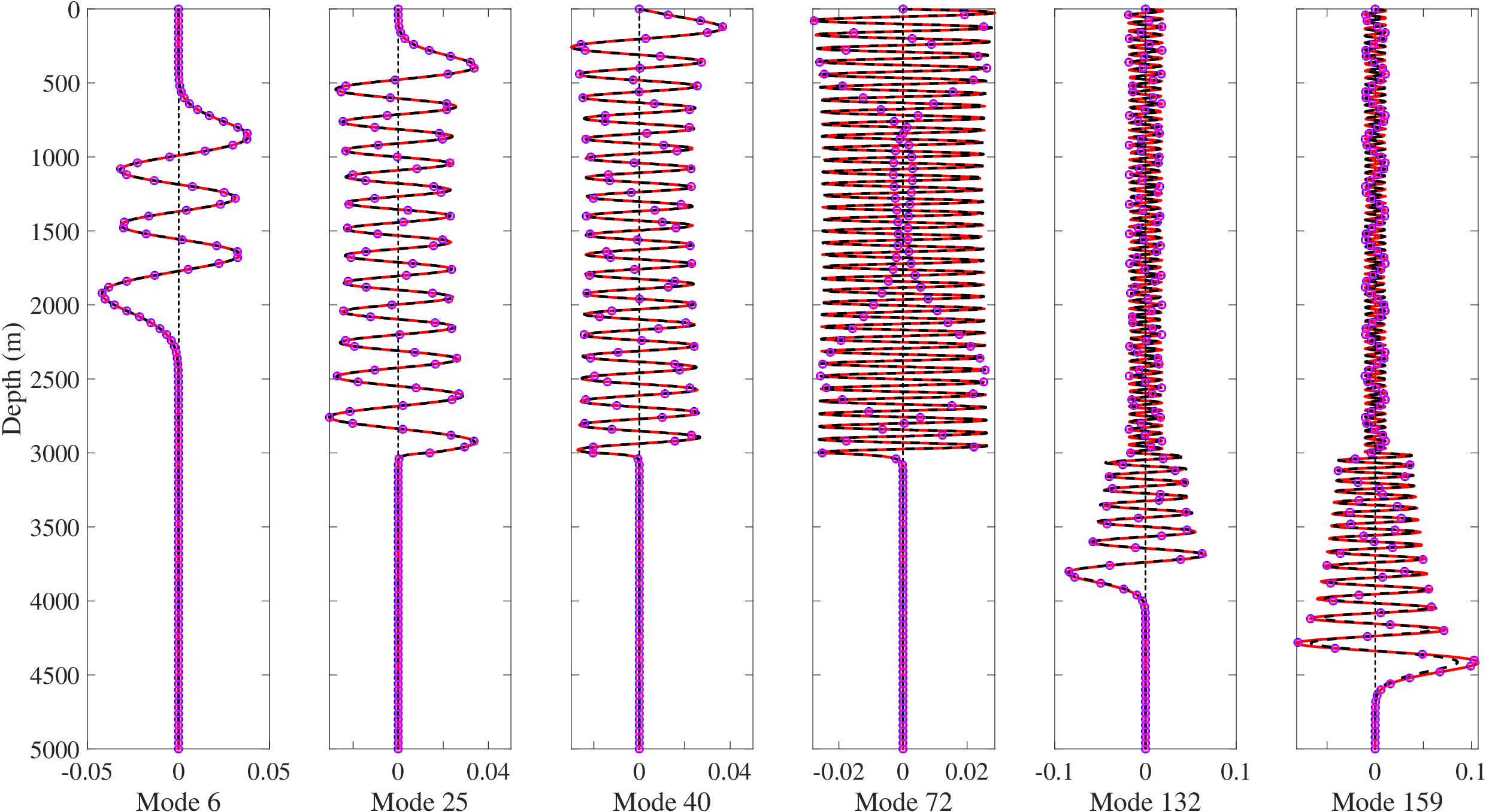}
		\caption{Mode shapes for Example 2 as calculated by Kraken (black dashed lines), rimLG and (small pink squares), NM-CT (small blue circles) and MultiLC (red solid lines) at 50 Hz.}
		\label{Figure4}
	\end{figure}

\begin{figure}[htbp]
		\centering
		\subfigure[]{\includegraphics[width=6.75cm]{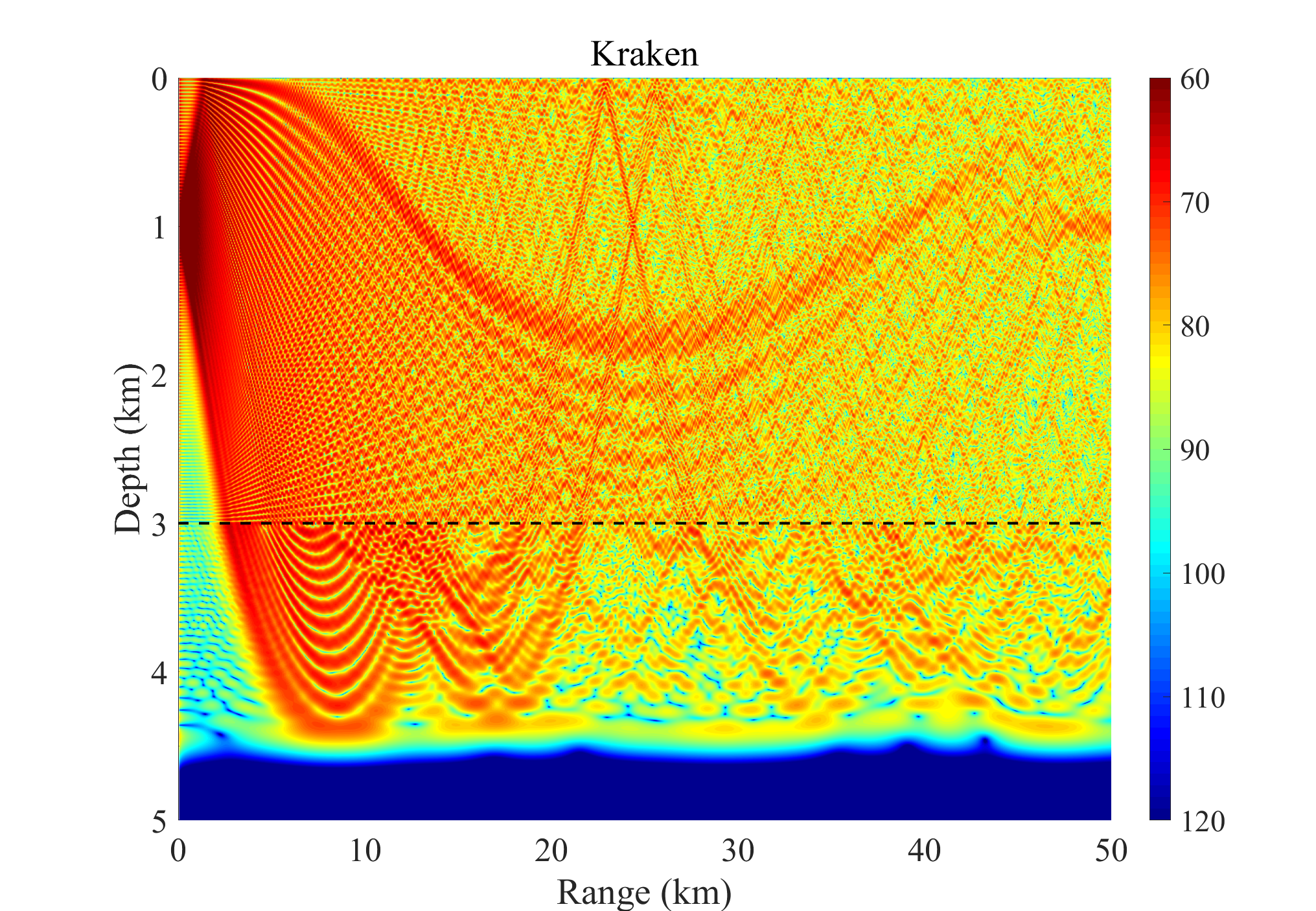}}
		\subfigure[]{\includegraphics[width=6.75cm]{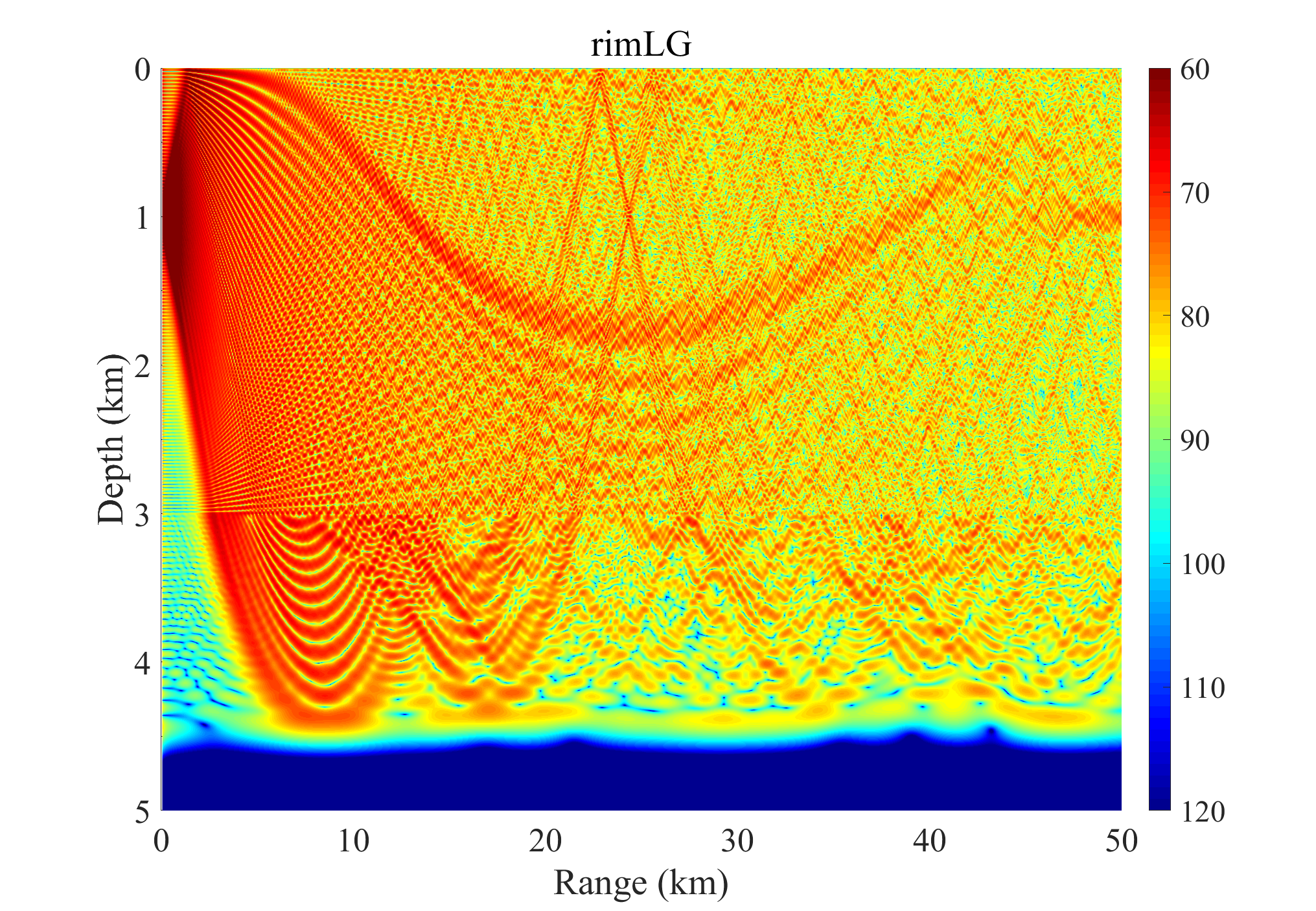}}
		\subfigure[]{\includegraphics[width=6.75cm]{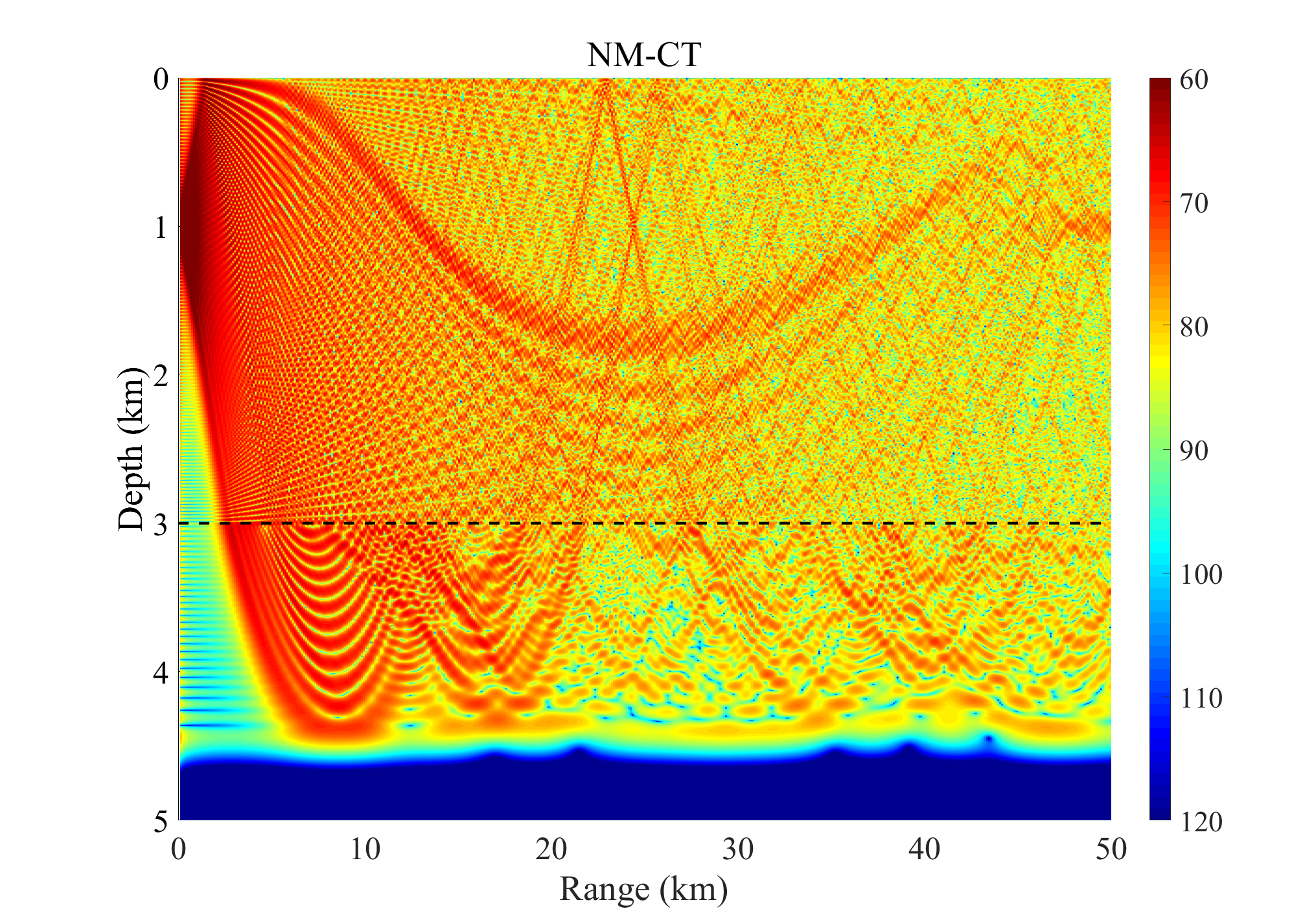}}
		\subfigure[]{\includegraphics[width=6.75cm]{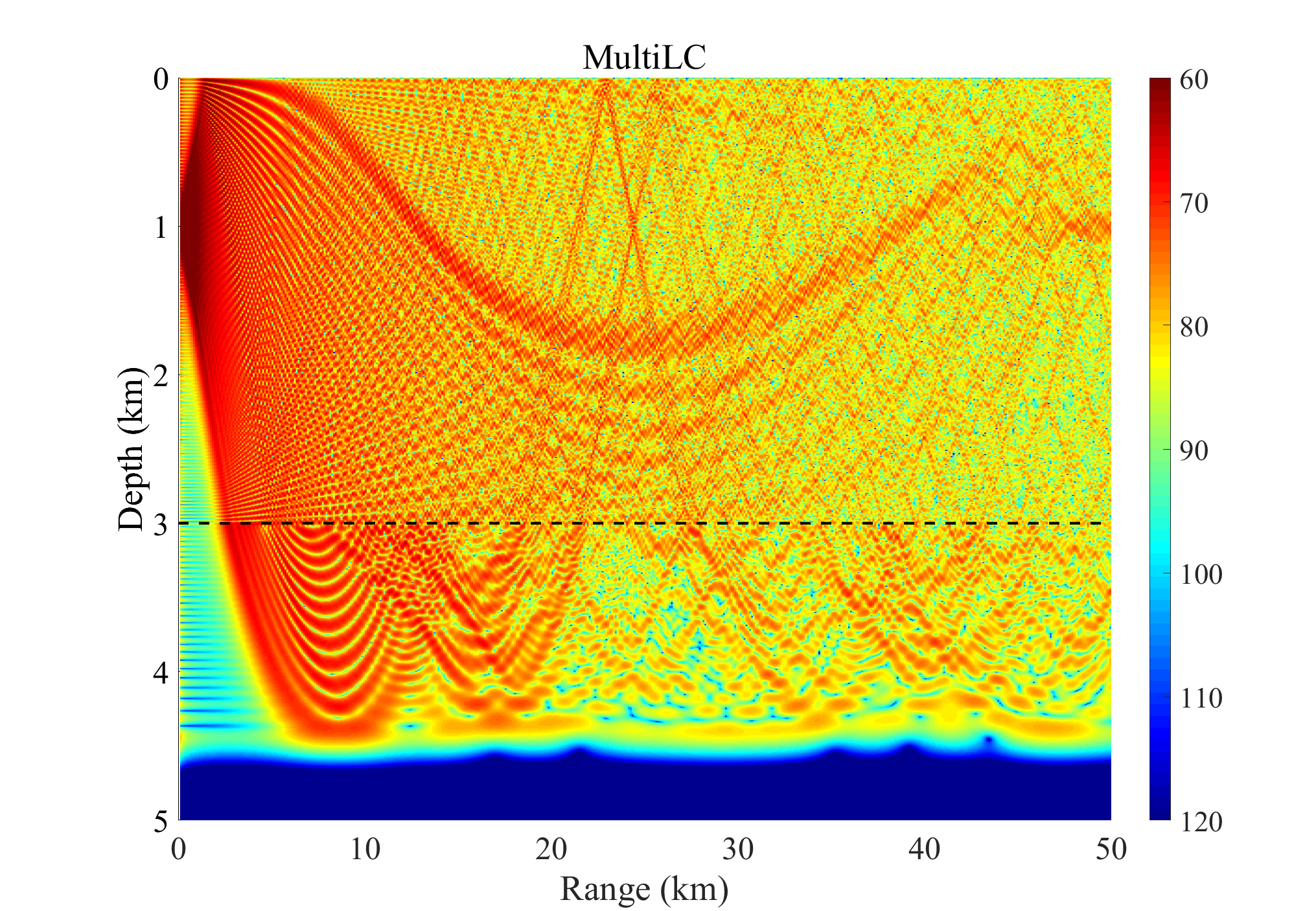}}
		\caption{Full-field TL results for Example 2 at 50 Hz as calculated by the 4 programs; the black dashed lines represent the interface between the water column and bottom sediment layer.}
		\label{Figure5}
	\end{figure}

	\begin{table}[htbp]
		\renewcommand\arraystretch{1.5}
		\scriptsize
		\caption{\label{tab4}Comparison of $k_{r_m}$ for Example 2.}
		\centering
		\begin{tabular}{cccccc}
			\hline
			{$f$ (Hz)}   & 
			{$m$  }      &
			{Kraken}	 &
			{rimLG}      &
			{NM-CT}      &
			{MultiLC}\\
			\hline
			50
			& 1      & 0.20937 35621 &0.20937 35621  &0.20937 35621 &0.20937 35621\\
			& 2      & 0.20924 24310 &0.20924 24310  &0.20924 24310 &0.20924 24310\\
			& 3      & 0.20911 22288 &0.20911 22288  &0.20911 22286 &0.20911 22288\\
			& 70     & 0.19485 68464 &0.19485 69442  &0.19485 69442 &0.19485 69442\\
			& 71     & 0.19446 46599 &0.19446 47584  &0.19446 47584 &0.19446 47584\\
			& 72     & 0.19406 59930 &0.19406 60923  &0.19406 60922 &0.19406 60923\\
			& 154    & 0.15931 10345 &0.15931 03136  &0.15931 03136 &0.15931 03136\\
			& 155    & 0.15882 12234 &0.15882 38594  &0.15882 38592 &0.15882 38593\\
			& 156    & 0.15860 37393 &0.15860 10722  &0.15860 10722 &0.15860 10722\\      
			\hline
			100
			& 1      & 0.41881 30249 &0.41881 30249  &0.41881 30249 &0.41881 30249\\
			& 2      & 0.41868 14240 &0.41868 14240  &0.41868 14239 &0.41868 14240\\
			& 3      & 0.41855 02932 &0.41855 02932  &0.41855 02932 &0.41855 02932\\
			& 150    & 0.38548 56958 &0.38548 57471  &0.38548 57470 &0.38548 57472\\
			& 151    & 0.38506 03554 &0.38506 04068  &0.38506 04066 &0.38506 04071\\
			& 152    & 0.38463 16785 &0.38463 17230  &0.38463 17297 &0.38463 17303\\
			& 311    & 0.31730 60645 &0.31730 53876  &0.31730 53876 &0.31730 53876\\
			& 312    & 0.31696 33628 &0.31696 65536  &0.31696 65535 &0.31696 65537\\
			& 313    & 0.31665 09930 &0.31664 48101  &0.31664 48100 &0.31664 48101\\  
			\hline
		\end{tabular}
	\end{table}
	Table \ref{tab4} shows the horizontal wavenumbers calculated by Kraken, NM-CT and MultiLC at frequencies of 50 Hz and 100 Hz. The truncation order of NM-CT and MultiLC is only 1000 at both frequencies, whereas Kraken uses 5000 discrete grid points in the vertical direction. The wavenumbers calculated by all three programs are very similar, with those of NM-CT and MultiLC being more similar than those of any other pairwise combination. 
	Fig.~\ref{Figure4} shows the shapes of six modes at 50 Hz for this example. The modes calculated by the three programs are very similar; however, there are slight deviations in the final mode. Fig.~\ref{Figure5} shows the TL fields at 50 Hz calculated by these three programs. The sound fields are again essentially indistinguishable.
	
\begin{figure}
		\subfigure{\includegraphics[width=6.75cm]{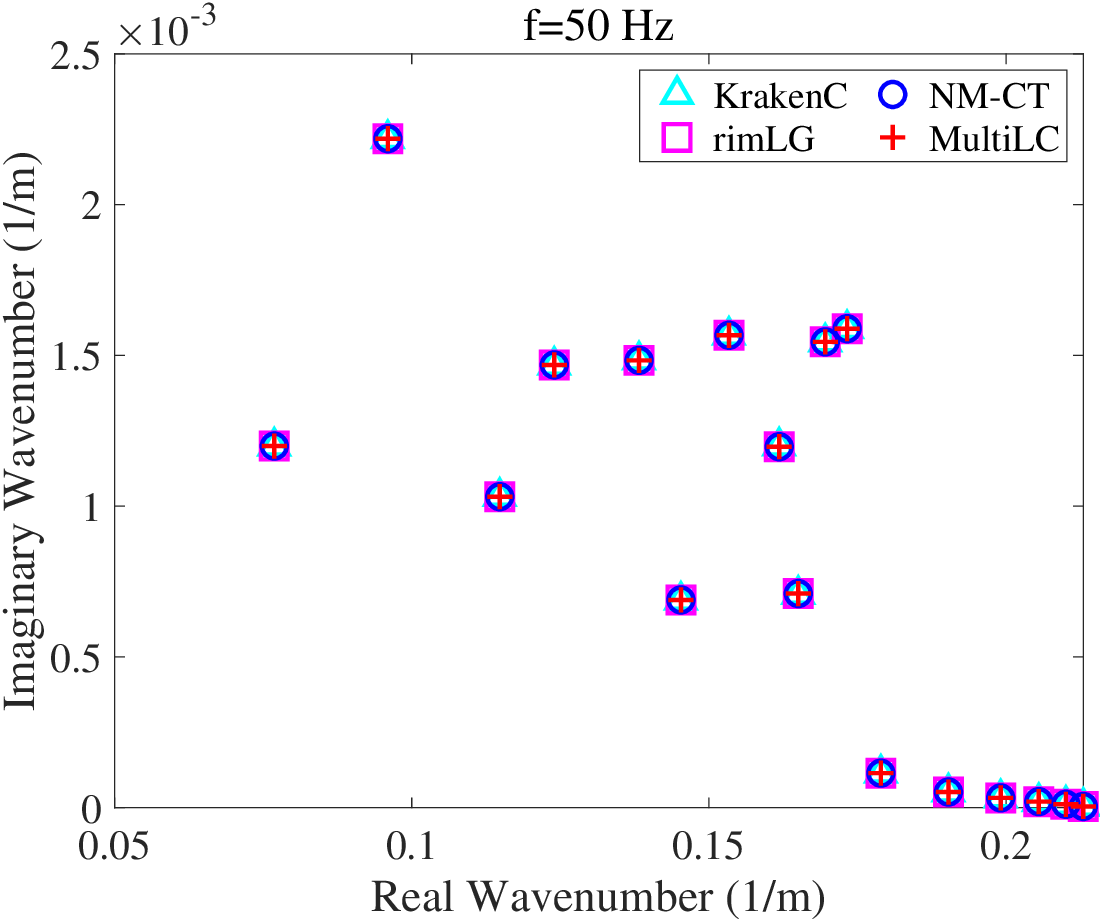}}
		\subfigure{\includegraphics[width=6.75cm]{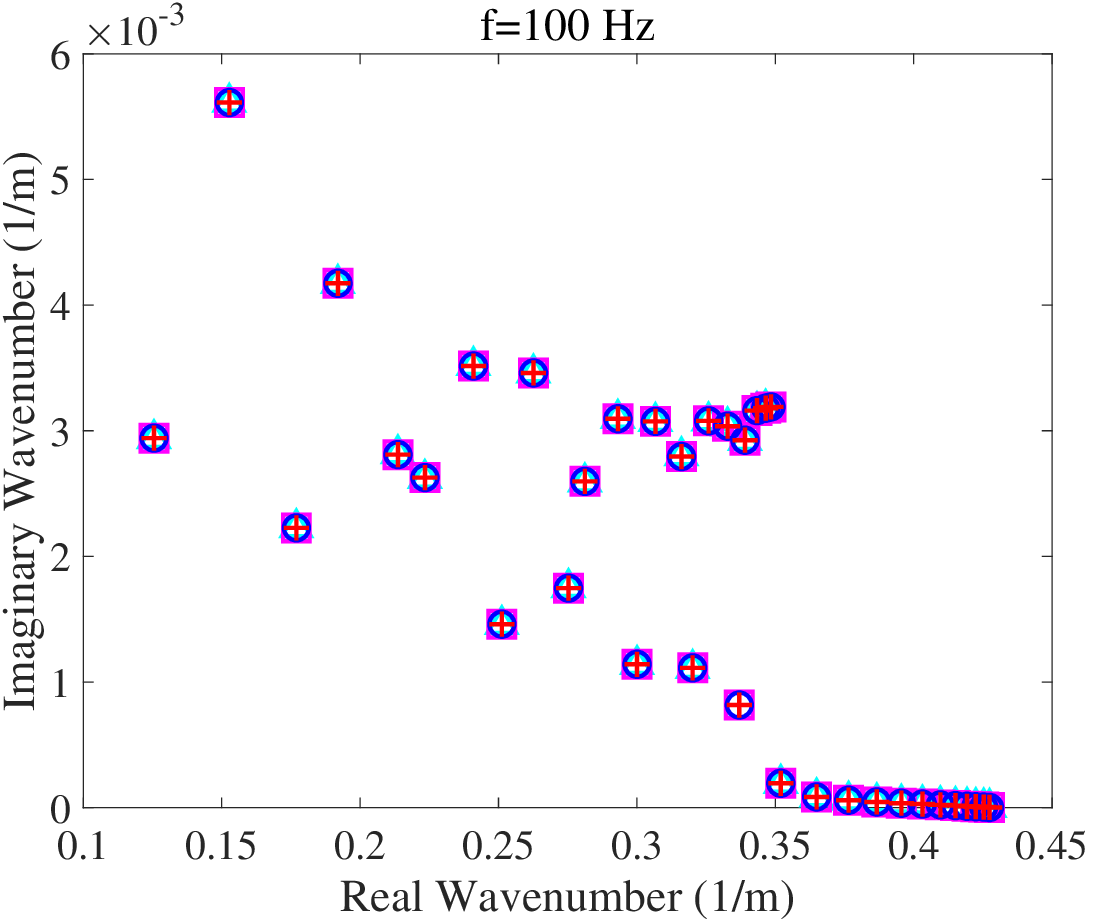}}
		\caption{Horizontal wavenumbers $k_{r_m}$ calculated by the 4 programs for Example 3.}
		\label{Figure6}
	\end{figure}

	Example 3 is a complex two-layer waveguide in which the sound speed has been measured. The previously presented derivation requires that the acoustic parameter profiles be sufficiently smooth; however, ocean acoustic parameter profiles that are measured in practice are not typically smooth. This example is presented to test the applicability of the LCM to actual ocean conditions. Fig.~\ref{Figure6} shows the horizontal wavenumbers calculated by all four programs at frequencies of 50 Hz and 100 Hz. The truncation orders of rimLG, NM-CT and MultiLC are only 100 at both frequencies, and the number of discrete points in the vertical direction used in KrakenC is automatically chosen by the code. As Fig.~\ref{Figure6} shows, the horizontal wavenumbers calculated by all programs are still similar. At $f$ = 50 Hz, the minimum difference in the horizontal wavenumber results between MultiLC and KrakenC is $3.617\times 10^{-7}+1.578\times 10^{-8}$i m$^{-1}$, and the maximum difference is $2.026\times 10^{-5} + 3.521\times 10^{-8}$i m$^{-1}$. At $f$ = 100 Hz, the minimum difference in the horizontal wavenumber results between MultiLC and KrakenC is $-1.594\times 10^{-7} + 1.050\times 10^{-7}$i m$^{-1}$, and the maximum difference is $2.104\times 10^{-4} - 4.515\times 10^{-4}$i m$^{-1}$. This accuracy shows that MultiLC also has the ability to handle acoustic parameter profiles that are not smooth. The run time comparison in Table \ref{tab5} yields the same conclusion as the previous examples---that MultiLC is the fastest of the three spectral methods. In this example, the sound speed profile is relatively rough, and the required truncation order of spectral methods is larger; thus, MultiLC is slightly slower than Kraken.
	
	\begin{table}[htbp]
		\renewcommand\arraystretch{1.5}
		\scriptsize
		\caption{\label{tab5}Comparison of the run times for Example 3 (unit: s).}
		\centering
		\begin{tabular}{cccccc}
			\hline
			$f$ (Hz) &KrakenC &MultiLC  &MultiLC (MATLAB)&rimLG (MATLAB) &NM-CT (MATLAB) \\ 
			\hline
			50   &0.133  &0.202  &0.047 &39.066 &0.262    \\
			100  &0.159  &0.228  &0.064 &42.382 &0.485    \\
			\hline
		\end{tabular}
	\end{table}
	
	\subsection{Semi-infinite space waveguide}
	Uniform semi-infinite space is a common boundary condition in the calculation of underwater acoustics. Sabatini et al. \cite{Sabatini2019} used the Chebyshev collocation method to discretize the governing equations into quadratic eigenvalue problems, which were mathematically solved perfectly. However, in actual calculations, the complexity of this solution process is relatively high. Equation~\eqref{eq:41} is the quadratic eigenvalue problem formed by Sabatini et al. \cite{Sabatini2019}. The equation shows that if the total number of collocation points is $N$, the square matrices $\boldsymbol{A},\boldsymbol{B},\boldsymbol{C}$ are also of order $N$, but the matrices $\boldsymbol{\tilde{A}},\boldsymbol{\tilde{B}}$ are of order $2N$, doubling the order of the matrices. 
	\begin{equation}
	    \label{eq:41}
        \begin{array}{c}
        \left[\boldsymbol{A}+k_{z, \infty} \boldsymbol{B}+k_{z, \infty}^{2} \boldsymbol{C}\right] \boldsymbol{\Psi}=\mathbf{0} \vspace{2ex}\\
        \boldsymbol{\tilde{A}} \boldsymbol{\tilde{\Psi}}=k_{z, \infty} \boldsymbol{\tilde{B}} \boldsymbol{\tilde{\Psi}} \vspace{2ex}\\
        \boldsymbol{\tilde{A}}=\left[\begin{array}{cc}
        -\boldsymbol{B} & -\boldsymbol{A} \\
        \boldsymbol{I} & 0
        \end{array}\right], \quad \boldsymbol{\tilde{B}}=\left[\begin{array}{cc}
        \boldsymbol{C} & 0 \\
        0 & \boldsymbol{I}
        \end{array}\right], \quad \quad \boldsymbol{\tilde{\Psi}}=\left[\begin{array}{c}
        k_{z, \infty} \boldsymbol{\Psi} \\
        \boldsymbol{\Psi}
        \end{array}\right]
        \end{array}
    \end{equation}
	When solving acoustic propagation in a complex ocean environment, this method is prone to produce a large-scale matrix that is difficult to solve. In addition, the solution of the quadratic eigenvalue problem is inherently unstable \cite{Francoise2000,Tisseur2001,Dedieu2003}. We use a different strategy to solve this problem to avoid costly calculations. As noted in Section~\ref{section2}, since the number of layers of the medium can be set arbitrarily, we absorb the downwardly propagating sound waves by setting a thick enough artificial absorbing layer below the depth of interest \cite{Evans2018,Beriot2020}; hence, a sound field similar to a uniform semi-infinite space can be obtained. Example 4 considers a simple Pekeris waveguide; the depth of the water column is 100 m, the density of seawater is 1 $\text{g}/\text{cm}^3$, and the speed of sound in the water is 1500 m/s. Below the water column is a uniform semi-infinite space with a density of 1.5 $\text{g}/\text{cm}^3$, a speed of sound of 2000 m/s, and an attenuation coefficient of 0.5 $\text{dB}/\lambda$. Assuming that we are concerned about the sound field above 100 m, the artificial absorbing layer with a thickness of 100 m and an attenuation coefficient of 5 $\text{dB}/\lambda$ can be set below 200 m. 
	
	\begin{table}[htbp]
		\renewcommand\arraystretch{1.5}
		\scriptsize
		\caption{\label{tab6}Comparison of $k_{r_m}$ for Example 4.}
		\centering
		\begin{tabular}{ccllll}
			\hline
			{$f$ (Hz)}   & 
			{$m$  }      &
			\makecell[c]{Analytical} &
			\makecell[c]{KrakenC}	 &
			\makecell[c]{Ref.~\cite{Sabatini2019}}      &
			\makecell[c]{MultiLC}\\
			\hline
			20
			&\multirow{2}{*}{1} 
			&0.07997 64228+  
			&0.07997 62711+   
			&0.07997 64344+      
			&0.07997 63941+ \\
			&
			&1.78360 6624e-5i
			&1.78326 3805e-5i 
			&1.78384 3798e-5i 
			&1.78347 90874e-5i \\
			&\multirow{2}{*}{2}
			&0.06709 82033+ 
			&0.06709 72484+   
			&0.06709 82118+ 
			&0.06708 82427+\\
			&  
			&0.00010 55733i
			&0.00010 55062i   
			&0.00010 55618i   
			&0.00010 43524i\\
			\hline
			50
			&\multirow{2}{*}{1}  
			&0.20750 84484+
			&0.20750 84277+    
			&0.20750 84674+ 
			&0.20750 84674+\\
			&
			&0.43350 55956e-5i
			&0.43431 25522e-5i 
			&0.43447 10290e-5i 
			&0.43451 03694e-5i\\
			&\multirow{2}{*}{2} 
			&0.20156 45436+
			&0.20156 44083+    
			&0.20156 45629+  
			&0.20156 45628+\\
			&  
			&0.16894 51109e-4i
			&0.16879 09257e-4i 
			&0.16885 13714e-4i
			&0.16886 66609e-4i\\
			&\multirow{2}{*}{3}
			&0.19115 74896+ 
			&0.19115 71581+    
			&0.19115 75099+   
			&0.19115 75098+ \\
			& 
			&0.38471 94422e-4i
			&0.38447 06435e-4i 
			&0.38461 23601e-4i
			&0.38464 71821e-4i \\
			&\multirow{2}{*}{4} 
			&0.17553 46439+
			&0.17553 39157+    
			&0.17553 46636+  
			&0.17553 46630+  \\ 
			&
			&0.82062 95415e-4i
			&0.82011 07025e-4i 
			&0.82049 82606e-4i   
			&0.82057 13814e-4i \\
			\hline
		\end{tabular}
	\end{table}
	
\begin{figure}
		\subfigure{\includegraphics[width=6.75cm]{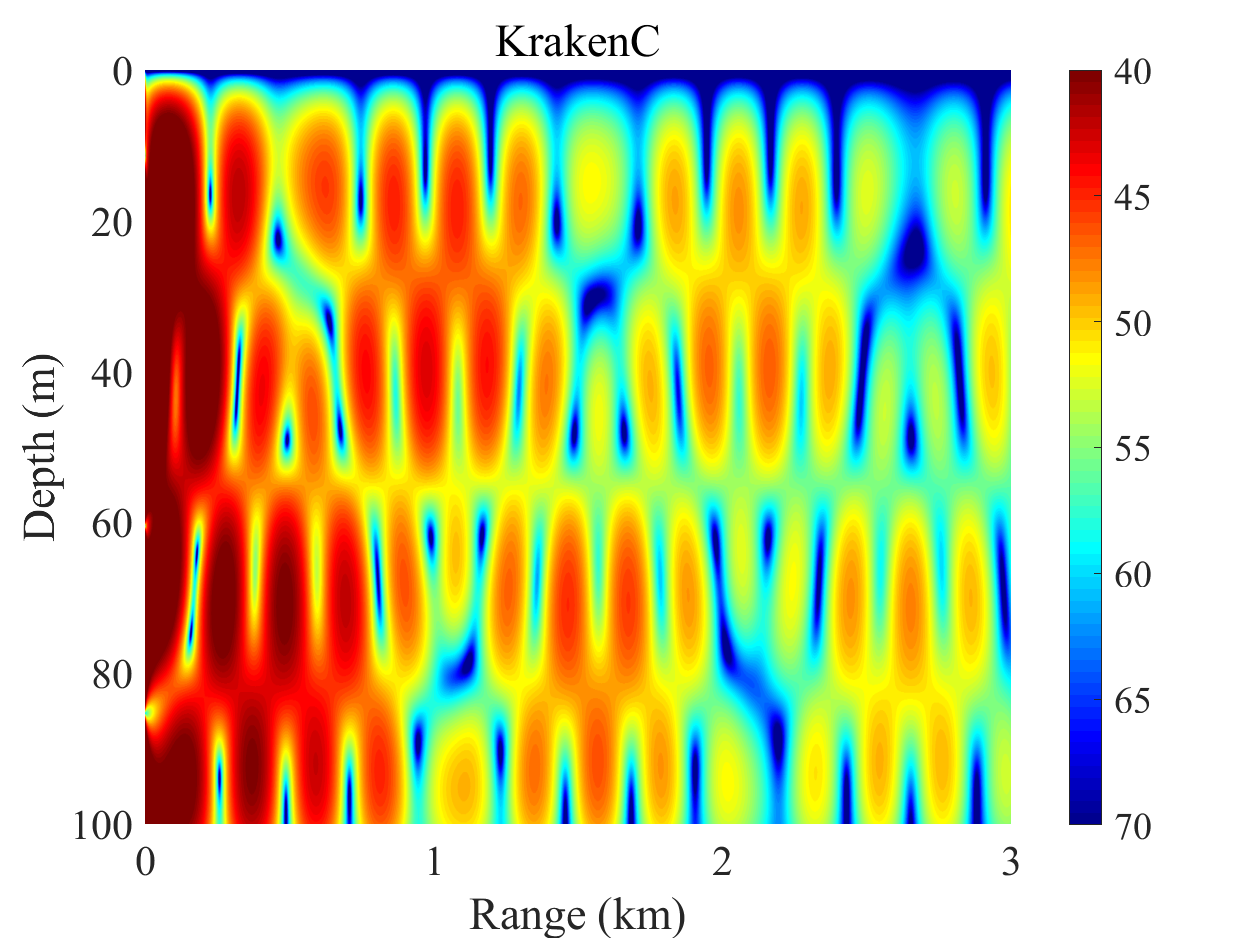}}
		\subfigure{\includegraphics[width=6.75cm]{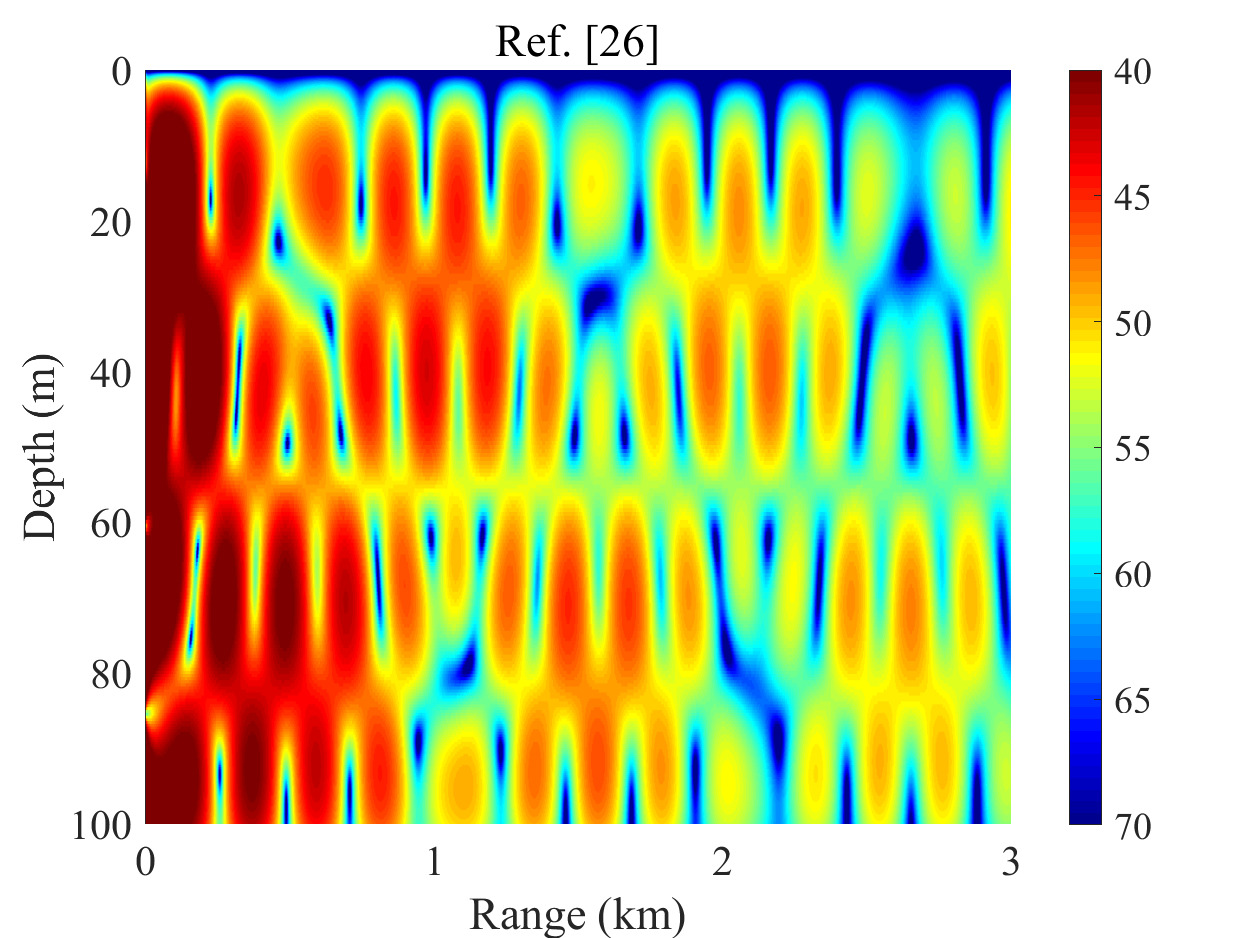}}
		\centering\subfigure{\includegraphics[width=6.75cm]{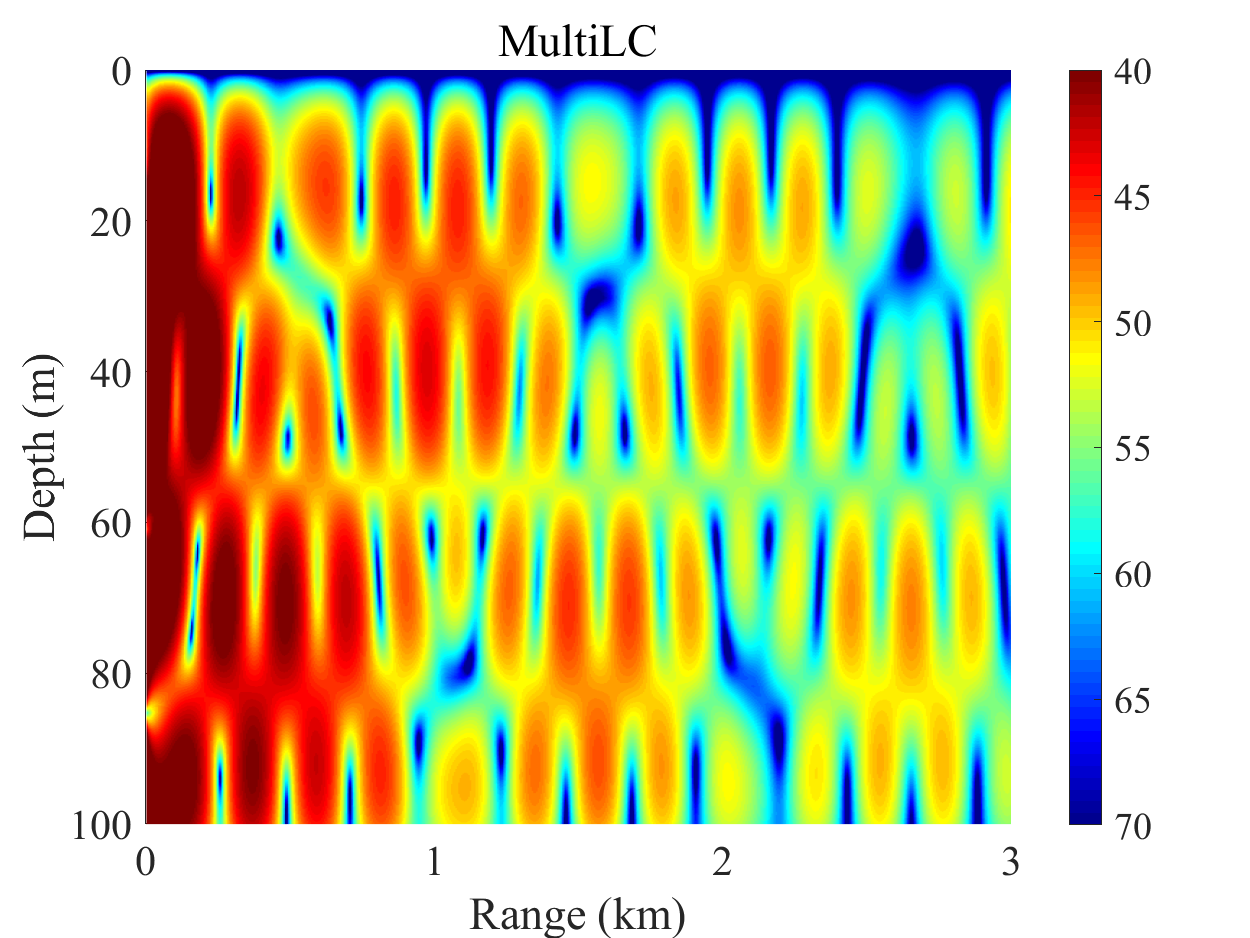}}
		\caption{Full-field TL results for Example 4 at 50 Hz as calculated by the 3 programs.}
		\label{Figure7}
	\end{figure}

	Table \ref{tab6} lists the horizontal wavenumbers of the Pekeris waveguide calculated by analytical solution \cite{Finn2011}, KrakenC, Sabatini \cite{Sabatini2019} and the MultiLC program developed in this paper. The table shows that the horizontal wavenumbers calculated by the three programs are very similar under the two source frequencies. Although there are still some small errors in the horizontal wavenumber, these errors have little effect on the final sound fields, as illustrated in Fig.~\ref{Figure7} (\textcolor{red}{The authors will add the sound field calculated by the analytical solution to Fig.~7 in the next revision.}).
	
	\section{Discussion}
	In the above examples, we verified the reliability of the MultiLC program when calculating simple ocean environments, layered oceans and semi-infinite subsea waveguides. The analysis of the above examples shows that in general, MultiLC requires fewer discrete points to achieve or exceed the accuracy of Kraken, especially when the acoustic profiles are smooth. When the acoustic profiles are not smooth, the spectral methods (rimLG, NM-CT and MultiLC) require a higher truncation order.
	
	For the semi-infinite subsea waveguides, we need to state the setting of the absorbing layer. The parameters of the artificial absorbing layer mainly include the thickness and the attenuation coefficient (the density and sound speed are the same as the half space). The thickness and the attenuation coefficient of the artificial absorbing layer influence each other because it is necessary to ensure that the current attenuation coefficient is sufficient to absorb the sound energy before it reaches the bottom of the layer in the current thickness of the layer. Therefore, the setting of the artificial absorbing layer (thickness and attenuation coefficient) relies on a certain degree of experience and tentativeness. 
According to the authors' experience, the thickness of the absorbing layer can be set to be similar to the thickness of other layers (so that a similar number of collocation points can be used and the calculation cost is acceptable). If the sound energy cannot be completely absorbed before reaching the bottom of the layer, the attenuation coefficient of the artificial absorbing layer should be increased (the absorption of sound energy can be judged from the transmission loss field diagram).
	
	Generally, compared with Kraken, MultiLC has higher accuracy; compared with other programs based on spectral methods (rimLG and NM-CT), the computational speed of MultiLC is much faster for the same accuracy. Compared with Kraken based on the finite difference method, MultiLC is faster when the marine acoustic profiles are smooth, while MultiLC is slightly slower than Kraken under a marine environment with rough acoustic profiles. For convenience of comparison, the same truncation order $N_i$ was used for the water columns and bottom sediment layers in MultiLC. In actual operation, the user can set $N_i$ arbitrarily according to the thickness of each layer and the complexity of the acoustic parameters. We have also tested the use of different $N_i$ values in the water columns and the bottom sediment layers; the results show that MultiLC can still generate sufficiently accurate horizontal wavenumbers and sound fields with such a configuration.
	
	\section{Conclusions}
	This article proposes a credible LCM that can solve the problem of horizontally stratified marine environmental sound propagation. The numerical experiments illustrate that compared to the program based on FDM, the program based on the LCM often requires fewer discrete grid points to achieve higher accuracy; this finding is sufficient to confirm that the LCM proposed in this paper is a high-precision algorithm. Compared with general spectral methods, LCM achieves approximately the same accuracy with the same configuration but consumes less time. Finally, in this article, the LCM was applied to simple range-independent cases only, and all the considered numerical examples are simple. Further research will be required to characterize the computational accuracy and speed of the LCM in a complex ocean environment. We anticipate that this study will help researchers gain a better understanding of how to solve the acoustic governing equations encountered in horizontally stratified oceans.
	
	\section*{Acknowledgments}
	The authors are very grateful to Roberto Sabatini for providing the program in Ref.~\cite{Sabatini2019}, and also very grateful to the two anonymous reviewers for their valuable comments.
	
	This work was supported by the National Key Research and Development Program of China [grant number 2016YFC1401800], the National Natural Science Foundation of China [grant numbers 61972406 and 51709267], and the Project of National University of Defense Technology [grant number 4345161111L].

\end{document}